\newif\ifcameraready
\camerareadytrue

\newif\ifdraft
\draftfalse

\newcounter{version}
\documentclass[letterpaper,nomarginnotes,nonarrowgutter]{jpaper}

\usepackage{cite}
\usepackage{amsmath,amssymb,amsfonts}
\usepackage{graphicx}
\usepackage{textcomp}
\usepackage{xcolor}
\usepackage{url}

\usepackage[colorinlistoftodos,prependcaption,textsize=scriptsize]{todonotes}

\usepackage{hyperref}

\usepackage{import}
\usepackage{algpseudocode}
\usepackage{amsmath}
\usepackage{multirow}
\usepackage{booktabs}
\usepackage{subcaption}
\usepackage{siunitx}
\usepackage{tabularx}
\usepackage{adjustbox}
\usepackage{bm}
\usepackage{enumitem}
\usepackage{multirow, makecell, amssymb}
\usepackage[normalem]{ulem}
\usepackage{tikz}
\usepackage{needspace}

\newcommand{\savefootnote}[2]{\footnote{\label{#1}#2}}
\newcommand{\repeatfootnote}[1]{\textsuperscript{\ref{#1}}}

\newcommand{\pudRef}{seshadri2013rowclone,seshadri2015fast,seshadri2016buddy,seshadri2016processing,seshadri2017ambit,li2017drisa,deng2018dracc,kim2012case,seshadri2019dram,deng2019lacc,wang2020figaro,xin2020elp2im,hajinazar2021simdram,ferreira2022pluto,wu2022dram,oliveira2022accelerating, deng2023dram,oliveira2024mimdram, prada, liu2025optipim, kubo2025mvdram, de2026count2multiply, olgun2022pidram, mutlu2024memory, mutlu2025memory, gao2022fracdram, kubo2025pudtune, oliveira2025proteus, li2025pim, gao2019computedram, tokuda2026clutch, afifi2024artemis,li2018scope, yaron2026bindram,garzon2026cadm,jahshan2024majork,ali2019memory,angizi2019graphide,bostanci2022drstrange,nadig2026conduit,
olgun2021quac,yuksel2023pulsar,yuksel2024functionally,yuksel2024simultaneous,yuksel2025pudhammer,kubo2024bulk,yukel2025simratrng,baser2026puf,luo2026dejavu,kim2018dramlatencypuf,kim2019drange} 

\newcommand{\pusdRef}{gao2019computedram, olgun2021quac, gao2022fracdram, yuksel2023pulsar, yuksel2024functionally, yuksel2024simultaneous, yuksel2025pudhammer, kubo2025pudtune, kubo2025mvdram, olgun2022pidram, mutlu2024memory, mutlu2025memory, yukel2025simratrng}
\newcommand{\pusdCite}{\cite{\pusdRef}}

\newcommand{\cotsdramCite}[0]{\cite{gao2019computedram,olgun2021quac,gao2022fracdram,yuksel2023pulsar,yuksel2024functionally,yuksel2024simultaneous,yuksel2025pudhammer,kubo2024bulk,kubo2025pudtune,kubo2025mvdram,olgun2022pidram, mutlu2024memory,mutlu2025memory,tokuda2026clutch, yauglikcci2022hira, yukel2025simratrng,yaron2026bindram,baser2026puf,luo2026dejavu}}

\newcommand{\pudappRef}
{hajinazar2021simdram, perach2023understanding, besta2021sisa, loving2014bitpal, wu2022dram, karunaratne2020memory, ali2019memory, angizi2019graphide, olgun2021quac, jahshan2024majork, garzon2026cadm, kubo2025mvdram, deng2023dram, de2026count2multiply, liu2025optipim,bostanci2022drstrange,
oliveira2025proteus, oliveira2024mimdram,seshadri2017ambit,li2025pim, tokuda2026clutch}
\newcommand{\pudappCite}{\cite{\pudappRef}}

\newcommand{\pudsimraRef}
{seshadri2017ambit, seshadri2015fast, hajinazar2021simdram, oliveira2024mimdram, de2026count2multiply, angizi2019graphide, jahshan2024majork,garzon2026cadm, liu2025optipim, ali2019memory, oliveira2025proteus, baser2026puf, gao2019computedram,olgun2021quac,gao2022fracdram,olgun2022pidram,yauglikcci2022hira, yuksel2024functionally,yuksel2024simultaneous, kubo2025pudtune, kubo2025mvdram, yukel2025simratrng,afifi2024artemis,yaron2026bindram,tokuda2026clutch}
\newcommand{\pudsimraCite}{\cite{\pudsimraRef}}

\newcommand{\pudCite}{\cite{\pudRef}}

\newcommand{\rhaRef}
{mutlu2017rowhammer,mutlu2019rowhammer,mutlu2023fundamentally,mutlu2023retrospective}

\newcommand{\isoAllCite}[0]
{\cite{
\rhaRef,
kim2014flipping,
luo2023rowpress,
yuksel2025columndisturb,
kim2020revisiting,
frigo2020trrespass,
olgun2025variable,
olgun2023hbm2rowhammer,
luo2025revisitingreaddisturbance,
tugrul2025reducedrefresh,
nam2024dramscope,
yaglikci2024spatialvariationaware,
yauglikcci2022understanding,
orosa2021deeperlook,
wang2026scaledisturb,
fournaris2017exploiting,
poddebniak2018attacking,
tatar2018throwhammer,
carre2018openssl,
barenghi2018software,
zhang2018triggering,
bhattacharya2018advanced,
seaborn2015exploiting,
safari2014rowhammer,
seaborn2015blackhat,
veen2016drammer,
gruss2016rowhammer,
razavi2016flip,
pessl2016drama,
xiao2016one,
bosman2016dedup,
bhattacharya2016curious,
burleson2016invited,
qiao2016new,
brasser2017cant,
jang2017sgx,
aga2017good,
tatar2018defeating,
gruss2018another,
lipp2018nethammer,
veen2018guardion,
frigo2018grand,
cojocar2019exploiting,
ji2019pinpoint,
hong2019terminal,
kwong2020rambleed,
cojocar2020susceptible,
weissman2020jackhammer,
zhang2020pthammer,
yao2020deephammer,
ridder2021smash,
jattke2022blacksmith,
tol2022toward,
kogler2022half,
orosa2022spyhammer,
zhang2022implicit,
liu2022generating,
cohen2022hammerscope,
zheng2022trojvit,
fahr2022frodo,
tobah2022spechammer,
rakin2022deepsteal,
khan2018analysis,
agarwal2018rowhammer,
li2014write,
ni2018write,
genssler2022reliability,
kakolyris2026columnkeeper,
kim2026pvac,
woo2026loaded,
park2016experiments,
luo2024combinedrowhammerrowpress,
ryu2017overcoming,
park2016statistical,
lim2017activeprecharge,
yun2018tid,
lim2018protonradiation,
lang2023blaster,
he2023whistleblower,
saxena2025citadel,
taneja2026mirza,
qureshi2026salt,
bostanci2025understanding,
chen2025rhohammer,
vittal2025mopac, 
woo2025mitigations, 
taneja2025dream,
chang2018voltron
}}

\newcommand{\dramdevRef}
{zhou2024understanding,demirel2025circuit, yang2019trap,zhou2023double,zhou2024unveiling,li2023understanding,walker2021dram,chang2018voltron,park2016experiments}
\newcommand{\dramdevCite}{\cite{\dramdevRef}}

\newcommand{\wallRef}
{wulf1995hitting, mutlu2019processing, dean2013tail, kanev2015profiling, mutlu2013memory, mutlu2014research, ferdman2012clearing, wang2014bigdatabench, mutlu2020intelligent, oliveira2021damov, boroumand2018google, boroumand2021google, wang2016reducing, koppula2019eden, ghose2019demystifying, mutlu2022modern, gomez2022benchmarking, kang2014co, mckee2004reflections, wilkes2001memory,wang2020figaro}
\newcommand{\wallCite}{\cite{\wallRef}}

\newcommand{\refthrCite}[0]{\cite{frigo2020trrespass,
ryu2017overcoming,
saxena2022aqua,
qureshi2022hydra,
marazzi2023rega,
lee2019twice,
hassan2021uncovering,
zhang2022softtrr,
marazzi2022protrr,
saroiu2024pracweb,
canpolat2025chronus,
canpolat2024securitybenefits,
kim2023ddr5prac,
jedec2024jesd795c,
saxena2024impress,
canpolat2024breakhammer,
bennett2021panopticon,
micron2016ddr4,
luo2023rowpress, 
yuksel2025columndisturb,
yuksel2025pudhammer,
yauglikcci2021blockhammer,
bostanci2024comet,
olgun2024abacus,
mutlu2023fundamentally,
kakolyris2026columnkeeper,
kim2026pvac,
woo2026loaded,
taneja2026mirza,
qureshi2026salt,
vittal2025mopac, 
woo2025mitigations, 
taneja2025dream
}}

\newcommand{\readDisturbRef}
{kim2014flipping,
park2016statistical,
park2016experiments,
lim2017activeprecharge,
ryu2017overcoming,
yun2018tid,
kim2020revisiting,
orosa2021deeperlook,
yauglikcci2022understanding,
luo2023rowpress,
luo2024combinedrowhammerrowpress,
yaglikci2024spatialvariationaware,
olgun2025variable,
tugrul2025reducedrefresh,
olgun2023hbm2rowhammer,
luo2025revisitingreaddisturbance,
lim2018protonradiation,
lang2023blaster,
nam2024dramscope,
he2023whistleblower,
luo2026dejavu,
wang2026scaledisturb,
yuksel2025pudhammer,
yuksel2025columndisturb,
chang2018voltron
}
\newcommand{\readDisturbCite}{\cite{\readDisturbRef}}

\newcommand{\openblRef}
{lim20121,
takahashi2001multigigabit,
chang2016low,
luo2020clr
}
\newcommand{\openblCite}{\cite{\openblRef}}

\newcommand{\PRE}{\texttt{PRE}}
\newcommand{\ACT}{\texttt{ACT}}

\newcommand{\WRITE}{\texttt{WRITE}}

\newcommand{\APA}{\texttt{APA}}

\newcommand{\AND}{\texttt{AND}}
\newcommand{\OR}{\texttt{OR}}

\newcommand{\RowCopy}{\texttt{RowCopy}}

\newcommand{\SiMRA}{\texttt{SiMRA}}

\newcommand{\MAJ}[1]{\texttt{MAJ#1}}
\newcommand{\MAJX}{\texttt{MAJX}}

\newcommand{\Frac}{\texttt{Frac}}

\newcommand{\blackcircled}[1]{%
  \tikz[baseline=(char.base)]{
    \node[shape=circle,fill=black,text=white,inner sep=1pt] (char) {#1};
  }%
}

\usepackage[most]{tcolorbox}
\tcbset{colframe=black!40!white,boxrule=0.5pt,sharp corners}
\newcounter{obsctr}
\newcounter{takectr}

\newtcolorbox{observation}[1][]{%
  enhanced, breakable,
  width=\dimexpr\linewidth\relax,
  colback=blue!12!white,
  left=1.5mm, right=1.5mm, top=0.6mm, bottom=0.6mm,
  before skip=2pt, after skip=2pt,
  fontupper=\normalsize,
  halign=left,
  before upper={%
    \refstepcounter{obsctr}%
    \noindent\textbf{Obsv.~\theobsctr.}
    \if\relax\detokenize{#1}\relax\else\ \textbf{#1}\fi\enspace
  }%
}

\newtcolorbox{stopbox}{%
  enhanced, breakable,
  width=\dimexpr\linewidth\relax,
  colback=red!12!white,
  left=1.5mm, right=1.5mm, top=2mm, bottom=2mm,
  before skip=2pt, after skip=2pt,
  fontupper=\Large\bfseries,
  halign=center,
}

\newtcolorbox{takeaway}[1][]{%
  enhanced, breakable,
  width=\dimexpr\linewidth\relax,
  colback=blue!12!white,
  left=1.5mm, right=1.5mm, top=0.6mm, bottom=0.6mm,
  before skip=2pt, after skip=2pt,
  fontupper=\normalsize,
  halign=left,
  before upper={%
    \refstepcounter{takectr}%
    \noindent\textbf{Takeaway~\thetakectr.}%
    \if\relax\detokenize{#1}\relax\else\ \textbf{#1}\fi\enspace
  }%
}

\newcommand{\numchips}{96}
\newcommand{\nummodules}{12}
\newcommand{\numobsv}{15}

\definecolor{gfored}{rgb}{0.580, 0.050, 0.211}
\definecolor{ao}{rgb}{0.007, 0.520, 0.867}
\definecolor{moegi}{rgb}{0.357, 0.537, 0.188}
\definecolor{jl}{rgb}{1.0, 0.2, 0.8}
\definecolor{brown(web)}{rgb}{0.65, 0.16, 0.16}
\definecolor{bisque}{rgb}{1.0, 0.89, 0.77}
\definecolor{nbs}{rgb}{0.88, 0.07, 0.37}
\definecolor{yt}{rgb}{0.58, 0.44, 0.86}
\definecolor{iy}{rgb}{0.0, 0.56, 0.041}
\definecolor{mel}{rgb}{0.9, 0.55, 0.31}
\definecolor{ouscolor}{rgb}{0.0, 0.2, 0.4}
\definecolor{dt}{rgb}{0.5, 0.1, 1.0}
\definecolor{jwangc}{rgb}{0.55, 0.1, 0.55}
\definecolor{agycolor}{rgb}{1.0, 0.2, 0.9}

\ifdraft
    \newcommand{\iey}[1]{\textcolor{iy}{#1}}
    
    \newcommand{\nb}[1]{\textcolor{nbs}{#1}}
    \newcommand{\jwang}[1]{\textcolor{jwangc}{#1}}
    \newcommand{\agyc}[1]{\textcolor{agycolor}{\textbf{[@gy:}~#1\textbf{]}}}
    \newcommand{\nbcomment}[1]{\textcolor{nbs}{\textbf{[@nb:}~#1\textbf{]}}}
    \newcommand{\jwangc}[1]{\textcolor{jwangc}{\textbf{[@jw:}~#1\textbf{]}}}
\else
    \newcommand{\iey}[1]{#1}
    
    \newcommand{\nb}[1]{#1}
    \newcommand{\agyc}[1]{}
    \newcommand{\nbcomment}[1]{}
    \newcommand{\jwang}[1]{#1}
    \newcommand{\jwangc}[1]{}
\fi

\definecolor{iyc}{rgb}{1.0, 0.0, 0.0}

\ifdraft
    \newcommand{\ieyc}[1]{\textcolor{iyc}{#1}}
\else
    \newcommand{\ieyc}[1]{}
\fi

\definecolor{dtokcol}{rgb}{0.05, 0.30, 0.85}
\ifdraft
  \newcommand{\dtok}[1]{\textcolor{dtokcol}{#1}}
\else
  \newcommand{\dtok}[1]{}
\fi

\definecolor{revA}{rgb}{0.580, 0.050, 0.211} 
\definecolor{revB}{rgb}{0.420, 0.460, 0.050}
\definecolor{revC}{rgb}{0.420, 0.280, 0.600} 
\definecolor{revD}{rgb}{0.550, 0.360, 0.090} 
\definecolor{revE}{rgb}{0.260, 0.420, 0.480} 
\definecolor{revF}{rgb}{0.000, 0.520, 0.200} 

\ifcameraready

    \newcommand{\atbcr}[2]{\ifnum#1>-1\textcolor{black}{#2}\else{#2}\fi}
    \newcommand{\ieycr}[2]{\ifnum#1>-1\textcolor{black}{#2}\else{#2}\fi}
    \newcommand{\omcr}[2]{\ifnum#1>-1\textcolor{black}{#2}\else{#2}\fi}
    \newcommand{\dtcr}[2]{\ifnum#1>-1\textcolor{black}{#2}\else{#2}\fi}
    \newcommand{\nbcr}[2]{\ifnum#1>-1\textcolor{black}{#2}\else{#2}\fi}
    \newcommand{\omcrcomment}[1]{}
    \newcommand{\crdiscussion}[2]{}{}

    \newcommand{\ominline}[1]{}
    \newcommand{\ieycrcomment}[1]{}
    \newcommand{\atbcrcomment}[1]{}
    \newcommand{\agycrcomment}[1]{}
    \newcommand{\dtcrcomment}[1]{}
    \newcommand{\ieyinline}[1]{}
    \newcommand{\dtinline}[1]{}
    \newcommand{\nbcrcomment}[1]{}
\else

\usepackage[colorinlistoftodos,prependcaption,textsize=scriptsize]{todonotes} 
    \paperwidth=\dimexpr \paperwidth + 4cm\relax
    \oddsidemargin=\dimexpr\oddsidemargin + 2cm\relax
    \evensidemargin=\dimexpr\evensidemargin + 2cm\relax
    \marginparwidth=\dimexpr \marginparwidth + 2cm\relax

    \newcommand{\atbcr}[2]{\ifnum#1=\value{version}\textcolor{ao}{#2}\else{#2}\fi}

    \newcommand{\ieycr}[2]{\ifnum#1=\value{version}\textcolor{blue}{#2}\else{#2}\fi}

    \newcommand{\nbcr}[2]{\ifnum#1=\value{version}\textcolor{nbs}{#2}\else{#2}\fi}

    \newcommand{\dtcr}[2]{\ifnum#1=\value{version}\textcolor{dt}{#2}\else{#2}\fi}

    \newcommand{\ieycrcomment}[1]{\todo[linecolor=orange, bordercolor=orange, backgroundcolor=white]{\textcolor{blue}{\textbf{@Ismail:} #1}}}
    \newcommand{\ieyinline}[1]{\\\textcolor{iy}{\textbf{@Ismail:} #1}}

    \newcommand{\nbcrcomment}[1]{\todo[linecolor=orange, bordercolor=orange, backgroundcolor=white]{\textcolor{nbs}{\textbf{@Nisa:} #1}}}

    \newcommand{\dtcrcomment}[1]{\todo[linecolor=orange, bordercolor=orange, backgroundcolor=white]{\textcolor{dt}{\textbf{@Daichi:} #1}}}
    \newcommand{\dtinline}[1]{\\\textcolor{dt}{\textbf{@Daichi:} #1}}

    \newcommand{\atbcrcomment}[1]{\todo[linecolor=brown, bordercolor=brown, backgroundcolor=white]{\textcolor{ao}{\textbf{@Atb:} #1}}}

    \newcommand{\agycrcomment}[1]{\todo[size=\scriptsize, linecolor=orange, bordercolor=orange, backgroundcolor=white]{\textcolor{gfored}{\textbf{@gy:} #1}}}
    
    \newcommand{\crdiscussion}[2]{\omcrcomment{#1\\\textcolor{dt}{\textbf{@Daichi:}#2}}}
    \newcommand{\omcr}[2]{\ifnum#1=\value{version}\textcolor{red}{#2}\else{#2}\fi}

    \newcommand{\omcrcomment}[1]{\todo[linecolor=orange, bordercolor=orange, backgroundcolor=white]{\textcolor{red}{\textbf{@Onur:} #1}}}
    \newcommand{\ominline}[1]{\\\textcolor{red}{\textbf{@Onur:} #1}}

\fi

\newcommand*\circledt[2]{\tikz[baseline=(char.base)]{
    \node[shape=circle, draw, fill=#1, inner sep=0.05pt] (char) {\vphantom{WAH1g}\textcolor{black}{#2}};}}

\hypersetup{
  colorlinks=false,
  pdfborder={0 0 0}
}

\newif\iffootnoterule

\makeatletter
\AtBeginDocument{%
\let\latex@@footnoterule\footnoterule

\renewcommand\footnoterule{%
  \iffootnoterule
  \latex@@footnoterule%
  \else
  \advance\skip\footins 4\p@\@plus2\p@\relax%
  \fi
}
}
\makeatother

\makeatletter
\g@addto@macro{\normalsize}{%
  \setlength{\abovedisplayskip}{4pt plus 0.5pt minus 1pt}
  \setlength{\belowdisplayskip}{3pt plus 0.5pt minus 1pt}
  \setlength{\abovedisplayshortskip}{0pt}
  \setlength{\belowdisplayshortskip}{0pt}
  \setlength{\intextsep}{5pt plus 1pt minus 1pt}
  \setlength{\textfloatsep}{3pt plus 1pt minus 1pt}
  \setlength{\skip\footins}{5pt plus 1pt minus 1pt}
  \setlength{\abovecaptionskip}{2pt plus 0pt minus 0pt}}
\makeatother

\usepackage{titlesec}
\titlespacing\section{0pt}{5pt plus 1pt minus 1pt}{2pt plus 1pt minus 1pt}
\titlespacing\subsection{0pt}{5pt plus 1pt minus 1pt}{2pt plus 1pt minus 1pt}
\titlespacing\subsubsection{0pt}{5pt plus 1pt minus 1pt}{2pt plus 1pt minus 1pt}

\newcommand{\thetitle}{PuDGhost: Experimental Analysis of Computation Result Corruption \\ in Processing-using-DRAM Operations on Real DRAM Chips \\ and Implications for Future Systems}
\title{\Large{\thetitle}\vspace{-0.4em}}

\author{
Daichi Tokuda\textsuperscript{1}\quad
{\.I}smail~Emir~Y{\"u}ksel\textsuperscript{2}\quad
Tatsuya Kubo\textsuperscript{1,4}\quad
Ataberk Olgun\textsuperscript{2}\quad
Haocong Luo\textsuperscript{2}\quad
Nisa Bostanci\textsuperscript{2}\\
Jikun Wang\textsuperscript{2}\quad
A.~Giray~Ya\u{g}l{\i}k\c{c}{\i}\textsuperscript{3}\quad
Shinya Takamaeda-Yamazaki\textsuperscript{1,4}\quad
Onur Mutlu\textsuperscript{2}\\[4pt]
\textsuperscript{1}The University of Tokyo\quad
\textsuperscript{2}ETH Zurich\quad
\textsuperscript{3}CISPA\quad
\textsuperscript{4}RIKEN
\vspace{-0.4em}
}

\begin{document}

\setstretch{0.955}
\maketitle

\makeatletter
\begingroup
  \renewcommand{\@makefnmark}{}
  \renewcommand{\@makefntext}[1]{\noindent\footnotesize #1}
\endgroup
\makeatother

\begin{abstract}
Processing-using-DRAM (PuD) is a promising computation paradigm to alleviate the frequent data movement between main memory and processing \dtcr{2}{units}.
The PuD paradigm provides a substrate for \dtcr{2}{highly} parallel computation by using each DRAM \dtcr{2}{column} as a computation engine via simultaneous multiple-row activation (\SiMRA{}).
Unfortunately, DRAM density scaling might hinder PuD's benefits. This is because denser cell arrays bring rows and columns closer, making even regular DRAM operations susceptible to noise and interference from neighboring cells. \dtcr{2}{PuD repurposes DRAM from a storage device into a parallel computing substrate, yet \textit{\emph{no}} prior work} investigates whether interference from rows or columns that are not intended to participate in the computation can compromise PuD robustness.

In this work, we reveal an interference phenomenon affecting PuD computations, \dtcr{2}{which we call} \emph{PuDGhost}, where a PuD operation \dtcr{2}{in a given column} produces erroneous results due to interference from 1)~data stored in non-activated DRAM rows and \dtcr{2}{2)~data stored in other columns that perform computations concurrently under the same \SiMRA{} operation.} 
PuDGhost violates the \dtcr{2}{ideal picture of PuD computations},
where each column's computation should depend solely on its \dtcr{3}{own} operand data. Thus, PuDGhost threatens the robustness of future PuD systems.
We present the first extensive characterization of PuDGhost using \numchips{} real DDR4 DRAM chips from \nummodules{} modules, systematically quantifying the impact of these two interference sources under various conditions \dtcr{2}{(i.e., data patterns, temperature, and spatial properties)}. \dtcr{2}{Among our \numobsv{} new empirical observations, we highlight two major results}: 1)~data in physically adjacent non-activated rows \dtcr{2}{affects} \SiMRA{} outputs by up to 10\% for random inputs, and \dtcr{2}{2)~data in columns that perform computations concurrently} \dtcr{2}{affects} \SiMRA{} outputs by up to 48\% for random inputs.
Guided by these findings, we propose countermeasures against PuDGhost across multiple layers of the PuD computing stack \dtcr{2}{(i.e., microarchitectural, architectural, and system levels)}. Specifically, we \dtcr{2}{propose and} evaluate on real \dtcr{2}{DDR4 DRAM} chips: 1) robust column screening that reduces the risk of mistakenly using unreliable columns in the presence of PuDGhost, and 2) a compute row layout that mitigates PuDGhost via \dtcr{2}{dedicated rows between compute rows}.
\dtcr{4}{Our solutions greatly improve PuD computation accuracy.}
We hope that our findings provide a foundation for developing solutions to enable future PuD systems \dtcr{2}{that are robust}.
\end{abstract}
\section{Introduction}
\dtcr{3}{Data} movement between main memory (DRAM) and processors has become a major bottleneck, \nb{consuming} a large share of execution time and energy in many \dtcr{3}{real workloads}~\wallCite{}. 
Processing-using-DRAM (PuD)~\pudCite{} is a promising paradigm that can alleviate this bottleneck by leveraging the existing \dtcr{3}{operational principles} of DRAM to realize massively parallel computation within DRAM. Prior work demonstrates
the potential of this approach to substantially improve throughput and energy efficiency compared to conventional systems for a wide range of applications, including databases, web search, data analytics, graph processing, genome analysis, cryptography, optimization solvers, hyperdimensional computing, and LLMs~\pudappCite{}.

\begin{figure}[htbp]
\centering
\includegraphics[width=\columnwidth]{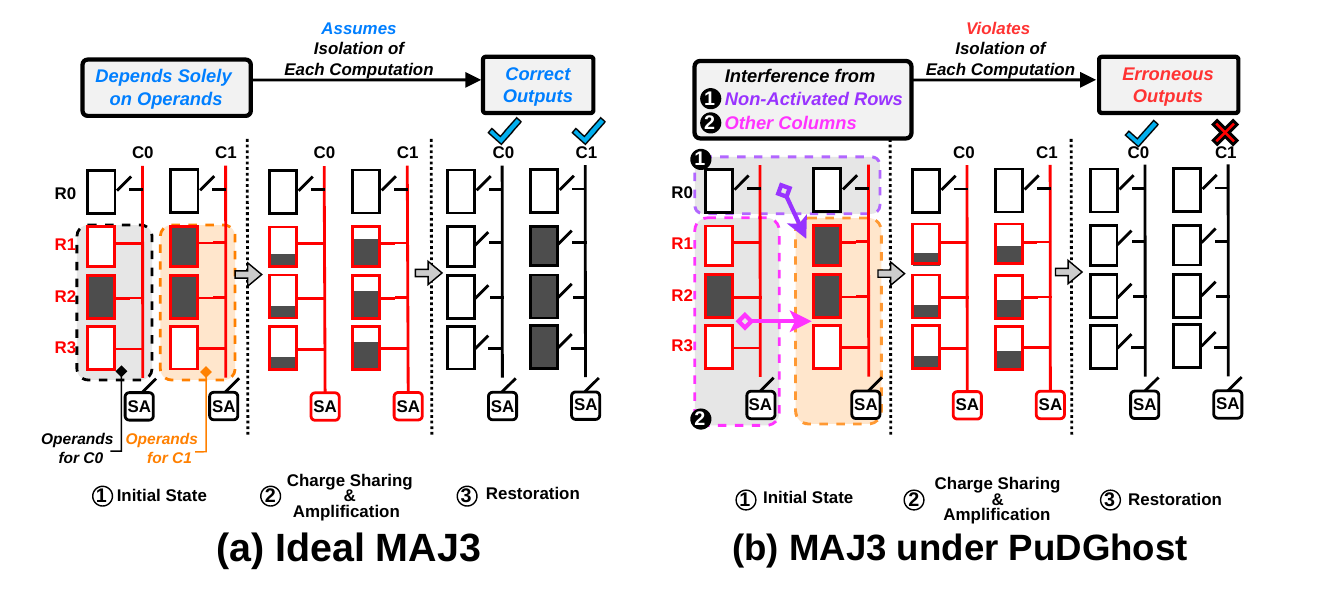}
\caption{(a) Ideal \MAJ{3}. (b) \MAJ{3} under PuDGhost.}
\label{fig:pudghost}
\end{figure}

The core computational \dtcr{3}{capability} of
various PuD architectures \dtcr{3}{relies} on Simultaneous Multiple-Row Activation (\SiMRA{}), a 
DRAM operation that simultaneously activates multiple DRAM rows within a subarray~\pudsimraCite{}. 
\ieycr{4}{Figure~\ref{fig:pudghost}a shows how performing \SiMRA{} \dtcr{3}{across} the cells of \ieycr{4}{simultaneously} activated \ieycr{4}{DRAM} rows (R1, R2, and R3) results in a majority operation (\MAJX{}) \dtcr{3}{to be computed} on the values stored in \dtcr{3}{the} multiple simultaneously activated \ieycr{4}{DRAM} cells in two \ieycr{4}{DRAM} columns (C0 and C1) in three key steps. First, the \MAJX{} input operands are initialized (\circledt{white}{1}). Second, the \SiMRA{} operation starts, enabling charge sharing among activated DRAM rows and columns (\circledt{white}{2}). Third, the \emph{sense amplifier} (SA in Figure~\ref{fig:pudghost}a) kicks in and samples the \MAJX{} result (\circledt{white}{3}) \dtcr{5}{based on its operational principles} (see \ref{sec:bac:pud}).}

\ieycr{4}{Ideally, as illustrated in Figure~\ref{fig:pudghost}a, the \MAJX{} result in each \ieycr{4}{DRAM} column is determined \emph{solely} by the activated cells used as operands. With each column performing its own \MAJX{}, each one of the many (e.g., 65536) DRAM columns \dtcr{3}{acts as} a computing \dtcr{3}{unit}, enabling the massive parallelism of PuD.}

\dtcr{3}{We hypothesize that} this ideal picture of PuD computations might be challenged by ongoing DRAM density scaling.
With rapid DRAM scaling and denser cell arrays,
even regular DRAM operations can
become susceptible to noise and interference from neighboring cells, \dtcr{3}{violating memory isolation (i.e., memory access to one location should not affect data stored in other memory locations)~\isoAllCite{}. Prior work~\cite{kim2014flipping, luo2023rowpress, yuksel2025columndisturb, kim2020revisiting, olgun2025variable,mutlu2019rowhammer} demonstrates that accessing one DRAM location can corrupt data in other locations, even in systems where DRAM is used \dtcr{5}{\emph{solely}} as a storage device.}
As PuD repurposes DRAM from a storage device to a \dtcr{5}{\emph{computation}} device, the notion of \emph{isolation} should extend beyond between memory accesses to each \emph{computation} (i.e., a computation's result should not be affected by data that is not intended to participate in the computation).
Unfortunately, to our knowledge, \emph{no} prior work investigates the impact of interference from rows or columns that are not intended to participate in the computation on \dtcr{3}{the reliability of PuD computation results}.

\dtcr{3}{This work is the first to reveal an interference phenomenon corrupting PuD computation results, which we call \textit{\textbf{PuDGhost}}, on real DRAM chips.
PuDGhost causes a PuD operation in a given column to produce erroneous results due to interference from non-operand data (i.e., data not intended to participate in the computation) stored in (1)~non-activated rows and (2)~other columns that perform computations concurrently under the same \SiMRA{} operation.}
\dtcr{3}{Figure~\ref{fig:pudghost}b illustrates how PuDGhost causes PuD computation errors. Rows R1, R2, and R3 are simultaneously activated to perform a \MAJ{3} operation. In column C1, the inputs are 1, 1, and 0, so the ideal \MAJ{3} output should be 1 (Figure~\ref{fig:pudghost}a). Due to PuDGhost, column C1 suffers interference from non-activated \dtcr{4}{rows} (e.g., R0) (\blackcircled{1}) and other columns (e.g., C0) (\blackcircled{2}), resulting in an erroneous output.}

We conduct the first extensive characterization of PuDGhost using \numchips{} real DDR4 DRAM chips (\nummodules{} modules). 
We systematically study interference from non-operand data under a wide range of operational conditions (e.g., data patterns, temperature, and spatial properties). 
Among our \numobsv{} key observations, we highlight two major findings.
First, non-operand data in non-activated rows that are physically adjacent to simultaneously activated rows affects \SiMRA{} outputs.
\dtcr{4}{Storing logic-0 (logic-1) in the adjacent rows biases the \SiMRA{} output toward logic-0 (logic-1), affecting \SiMRA{} outputs by up to 10\% for random inputs. The bias increases monotonically as the fraction of logic-1 in adjacent rows increases (detailed in \S\ref{sec:row:adj}).}
Second, non-operand data in other columns that concurrently perform computations under the same \SiMRA{} operation (i.e., the inputs of these concurrent computations) \dtcr{4}{also affects \SiMRA{} outputs by up to 48\% for random inputs. Unlike adjacent-row interference, column-wise interference is both stronger and exhibits a non-monotonic relationship with the fraction of logic-1 in these columns' inputs (detailed in \S\ref{sec:col}).}

Our \dtcr{3}{real DRAM chip} characterization results suggest that PuDGhost is an important consideration for \dtcr{3}{designing} future PuD systems that are robust.
Building on \dtcr{3}{our empirical} insights, we analyze PuDGhost's impact on PuD reliability and present \dtcr{3}{robust PuD} solution directions across multiple layers of the PuD computing stack. 
At the system level, 
\dtcr{4}{we reveal that PuD systems that are unaware of adjacent-row data during column screening and PuD execution can mislabel unreliable columns as reliable.}
\dtcr{4}{We propose robust column screening methods that control the data in rows adjacent to compute rows during both screening and PuD execution, reducing the risk of mistakenly using unreliable columns in the presence of PuDGhost.}
At the architecture level, we propose a compute row layout that uses dedicated isolation rows with fixed data patterns between compute rows, \dtcr{4}{ensuring that the rows adjacent to compute rows store fixed data that does not change during PuD execution.}

We evaluate \dtcr{5}{our} solutions on real DRAM chips \dtcr{5}{for} two use cases: general matrix-vector multiplication (GEMV) and true random number generation (TRNG). \dtcr{4}{Our results demonstrate that our solutions significantly reduce the impact of PuDGhost, providing 1)~413× lower normalized mean squared error (NMSE) in GEMV and 2)~preventing 93\% of the entropy loss in TRNG, compared to PuDGhost-unaware systems.}

Our main contributions are as follows.
\begin{itemize}
\item \dtcr{3}{We perform the first experimental study of a \dtcr{4}{new} interference phenomenon \dtcr{4}{in DRAM} \dtcr{4}{that corrupts} Processing-using-DRAM (PuD) computation results, which we call PuDGhost, on real DRAM chips.}

\item \dtcr{3}{Our experimental results on real DDR4 DRAM chips} reveal that PuD computation results in a given column can be affected by data stored in 1)~non-activated rows adjacent to the rows used for computation and 2)~other columns that perform computations concurrently. PuDGhost violates the \dtcr{4}{expectation} \dtcr{3}{that each computation depends solely on its own operand data.}

\item We propose and discuss solution\dtcr{4}{s} across multiple layers of the PuD computing stack to reduce PuDGhost-induced \dtcr{3}{PuD computation} errors. 

\item We evaluate our solutions on real DRAM chips \dtcr{5}{for} two use cases: general matrix-vector multiplication (GEMV) and true random number generation (TRNG). Our results demonstrate that our solutions significantly reduce the impact of PuDGhost \dtcr{5}{on these real use cases}.

\item We believe and hope that our results and analyses will enable and inspire future research to reduce computation errors in PuD, and to design future PuD systems that are robust.
\end{itemize}
\footnoteruletrue
\section{Background}
\subsection{DRAM Organization and Operation}
\begin{figure}[h]
\centering
\includegraphics[width=0.90\columnwidth]{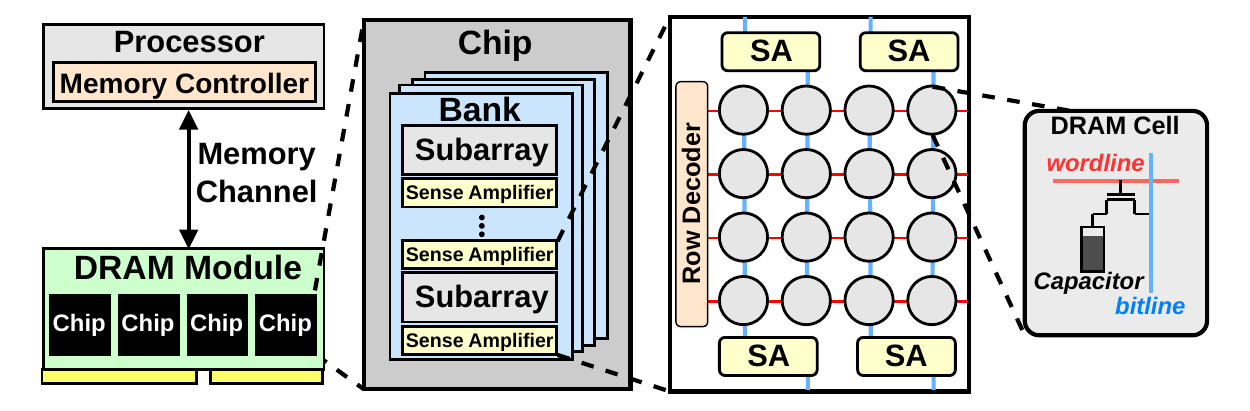}
\caption{DRAM Organization.}
\label{fig:dram}
\end{figure}

Dynamic Random Access Memory (DRAM) is organized in a hierarchical structure consisting of channels, ranks, chips, banks, and subarrays of memory cells (Figure~\ref{fig:dram}). A module contains one or more ranks, and each rank consists of multiple DRAM chips. Each chip has multiple banks (e.g., 8–16), and each bank is further divided into multiple subarrays.
Within each subarray, DRAM cells form a two‑dimensional grid of rows (wordlines) and columns (bitlines).
Each cell consists of a single transistor paired with a capacitor, and stores one bit of data, based on the charge level held in the capacitor. 
The DRAM cells in the same column are connected to the sense amplifier (SA) via a bitline. 
Modern DRAM employs an open bitline architecture~\openblCite{}, where half of the bitlines in a subarray share SAs with the upper adjacent subarray and the other half share SAs with the lower adjacent subarray.
The memory controller integrated in the CPU die generates a sequence of DRAM commands to access data in DRAM. The \ACT{} command opens a specific row and copies its data into the row buffer. The \PRE{} command closes the active row. These commands operate on all columns in a row. 

\subsection{Processing-using-DRAM}\label{sec:bac:pud}
\noindent
\textbf{Computational Capability of PuD.}
Processing-using-DRAM (PuD)~\pudCite{} is a paradigm that can alleviate the bottleneck caused by frequent data movement between processing elements (e.g., CPUs) and main memory. PuD enables massively parallel computation within DRAM by leveraging the intrinsic analog operational properties of DRAM circuitry.
Many PuD architectures perform computation through two \dtcr{4}{primitives}: 1) in-DRAM data copy from one row to another (\RowCopy{}) using consecutive multiple-row activation~\cite{seshadri2013rowclone, seshadri2017ambit, seshadri2015fast,seshadri2019dram, gao2019computedram, olgun2022pidram, yuksel2024simultaneous}, and 2) in-DRAM bitwise operations using simultaneous multiple-row activation (\SiMRA{})~\pudsimraCite.
In bitwise operations using \SiMRA{}, simultaneously activating multiple DRAM rows within the same subarray induces charge sharing \dtcr{4}{of activated cells in each bitline}, \dtcr{4}{resulting in} a majority-of-X operation (\MAJX{}) in each column. For example, when three rows are simultaneously activated, the charges of cells on the same bitline combine through charge sharing to produce the \MAJ{3} result (see Figure~\ref{fig:pudghost}a). \MAJX{} can implement basic Boolean operations such as \AND{} and \OR{}, and by chaining multiple \MAJX{} operations, PuD can accelerate a wide range of computations, from basic arithmetic to complex kernels including general matrix-vector multiplication (GEMV) for LLM inference~\pudappCite{}.

In Ambit~\cite{seshadri2017ambit, seshadri2015fast,seshadri2019dram} and its successor architectures~\cite{hajinazar2021simdram, oliveira2024mimdram, oliveira2025proteus}, six \emph{compute rows} per subarray are reserved to execute \MAJ3{}. The row decoder is modified to allow specific triplet combinations of these six rows to be simultaneously activated, enabling a \MAJ3{} operation. The remaining rows in the same subarray serve as storage rows, holding data that is loaded into compute rows via \RowCopy{} when needed for computation.

\noindent
\textbf{PuD on COTS DRAM Chips.}
Prior work~\cotsdramCite{} demonstrates that commercial off-the-shelf (COTS) DRAM chips possess PuD computation capability.
Specifically, prior work \cite{yuksel2024functionally, yuksel2024simultaneous, yuksel2025pudhammer, yukel2025simratrng, yuksel2023pulsar} experimentally shows that COTS DDR4 chips from SK Hynix~\cite{skhynix_ddr4} can simultaneously activate 2, 4, 8, 16, or 32 rows within a subarray by violating nominal timing parameters. \dtcr{4}{The memory controller can perform} \SiMRA{} \dtcr{4}{on these chips} by issuing an \ACT{}-\PRE{}-\ACT{} \dtcr{4}{command sequence} (\APA{} sequence) with very short intervals of 3ns or less \dtcr{5}{between each command}.

\noindent
\textbf{PuD Computation Errors.}
Prior work focuses primarily on process variation in DRAM circuit components as the mechanism for \MAJX{} errors.
Prior work~\cite{seshadri2017ambit, yuksel2024simultaneous,hajinazar2021simdram,yuksel2023pulsar} shows through circuit-level simulations that \MAJ{3} operations produce erroneous results when process variation of subarray components (e.g., cell capacitance) is large. 
Characterization studies using COTS DRAM chips~\cite{yuksel2023pulsar,yuksel2024functionally, yuksel2024simultaneous, kubo2025pudtune,gao2019computedram,gao2022fracdram} experimentally show that the success rate of \MAJX{} on real DRAM chips is \jwang{below} 100\%, and that error susceptibility varies across columns. 
In these prior studies, the causes of \SiMRA{}-based \MAJX{} errors have been attributed to process variation of cell capacitance, access transistors, and SAs in each column. 

\section{Methodology}
We describe our COTS DRAM chip testing infrastructure (\S\ref{sec:met:inf})
and the COTS DDR4 chips tested for our characterization study
(\S\ref{sec:met:chi}). 

\subsection{COTS DRAM Testing Infrastructure}\label{sec:met:inf}
\begin{figure}[h]
    \centering
    \includegraphics[width=0.60\columnwidth]{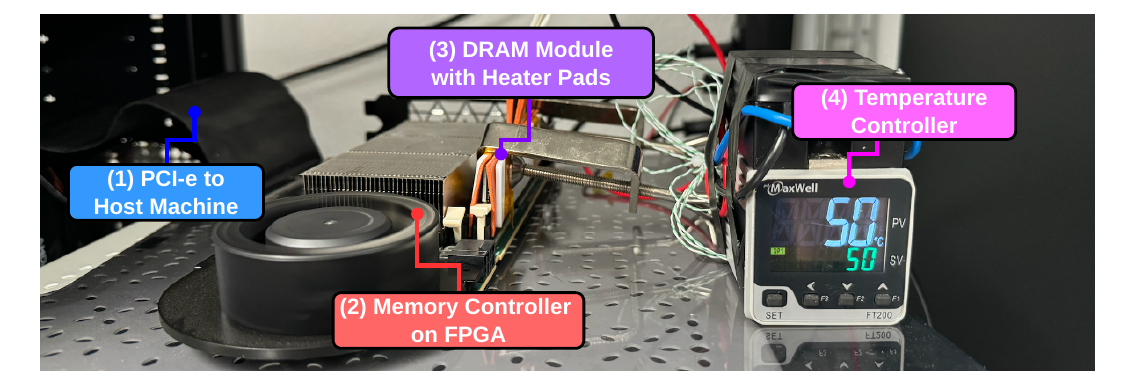}
    \caption{Our FPGA-based PuD testing infrastructure (DRAM Bender~\cite{olgun2023dram,safari2022drambender}) with DDR4 modules.}
    \label{fig:infra}
\end{figure}

We conduct COTS DRAM chip experiments using DRAM Bender~\cite{olgun2023dram,safari2022drambender, hassan2017softmc,safari2017softmc}, an FPGA-based DDR4
testing infrastructure that provides precise control of DDR4 commands.
Figure~\ref{fig:infra} shows our experimental setup that consists of four
main components: 1) a host machine that generates the test program
and collects results, 2) an FPGA development board~\cite{xilinx_alveo_u200},
programmed with DRAM Bender, 3) thermocouple temperature
sensors and heater pads pressed against the DRAM chips to maintain
target temperature levels, and 4) a temperature controller
that keeps the temperature at the desired level.

\subsection{COTS DDR4 DRAM Chips Tested}~\label{sec:met:chi}
Table~\ref{tab:dram} lists the 96 COTS DDR4 DRAM chips \iey{from 12 modules}, showing chip manufacturer (Chip Mfr.),~\jwang{module manufacturer (Module Mfr.),} module count (\#Modules), chip count (\#Chips), die revision (Die Rev.), density, and \jwang{chip} organization (Org.). All tested chips are from SK Hynix, as prior \dtcr{4}{work reports} that only SK Hynix modules can perform \SiMRA{}~\pusdCite{}.
We test modules from various module manufacturers, die revisions, and chip densities so that our findings apply across different DRAM designs and manufacturing processes.\footnote{\dtcr{4}{Prior work~\cite{yuksel2024simultaneous, yuksel2023pulsar,olgun2021quac,yuksel2024functionally} hypothesizes that the hierarchical row decoder design is the primary enabler of \SiMRA{} on COTS DRAM chips: reducing the intervals between \APA{} sequence allows the local wordline decoder to latch the subsequent row address without de-asserting the previous one. 
We believe that chips from other vendors are also fundamentally capable of \SiMRA{}, as \SiMRA{} leverages the hierarchical row decoder design that is common across high-performance DRAM chips and is likely to persist in future generations.}}

\begin{table}[h]
  \centering
  \caption{Summary of DDR4 DRAM chips tested.}
  \label{tab:dram}
  \setlength{\tabcolsep}{5pt}
  \renewcommand{\arraystretch}{1.05}
  \begin{tabular}{@{}l l c c c c@{}}
    \toprule
    \makecell{\textbf{Chip}\\\textbf{Mfr.}} &
    \makecell{\textbf{Module}\\\textbf{Mfr.}} &
    \makecell{\textbf{\#Modules}\\\textbf{(\#Chips)}} &
    \makecell{\textbf{Die}\\\textbf{Rev.}} &
    \makecell{\textbf{Chip}\\\textbf{Density}} &
    \makecell{\textbf{Chip}\\\textbf{Org.}} \\
    \midrule\midrule
    \multirow{3}{*}{SK Hynix} & TimeTec   & 3 (24) & A & 4Gb & x8 \\
                              & TeamGroup & 7 (56) & M & 4Gb & x8 \\
                              & SK Hynix  & 2 (16) & A & 8Gb & x8 \\
    \bottomrule
  \end{tabular}
\end{table}

\noindent
\textbf{Logical-to-Physical Row Mapping.}
DRAM manufacturers use mapping schemes to translate logical to physical row addresses. To account for in-DRAM row address mapping, we reverse engineer the physical row address layout in all tested chips by analyzing RowHammer-induced bitflip patterns, following prior methodologies~\cite{kim2020revisiting, luo2023rowpress, yuksel2024simultaneous,yuksel2024functionally,yuksel2025pudhammer}.

\noindent
\textbf{Subarray Boundaries.}
Following prior methodologies~\cite{yuksel2024simultaneous, yuksel2024functionally, yuksel2025pudhammer, nam2024dramscope,gao2019computedram}, we identify subarray boundaries using \RowCopy{}, which only succeeds when source and destination rows are in the same subarray. By attempting \RowCopy{} across consecutive row pairs, we reconstruct each chip’s subarray map.

\noindent
\textbf{True/Anti Cells.}
DRAM cells are classified as true-cell or anti-cell based on how a fully charged capacitor is interpreted: in a true-cell (anti-cell), a charged capacitor represents logic-1 (logic-0) and a discharged capacitor represents logic-0 (logic-1). Prior work~\cite{liu2013experimental, nam2024dramscope} on DRAM retention failure commonly assumes that retention-induced errors are from the charged to discharged state.
We identify the cell type of our chips following these works. Throughout this paper, logic-1 denotes a charged capacitor and logic-0 denotes a discharged capacitor.

\noindent
\textbf{Even/Odd Columns.}
In an open-bitline layout~\openblCite{}, adjacent subarrays share sense amplifiers (SAs). Bitlines alternate their connections: even columns connect to SAs on one side while odd columns connect to SAs on the other side. We identify this even/odd column assignment by analyzing \RowCopy{} outcomes and charge-sharing across subarray boundaries, following prior work~\cite{yuksel2024functionally,nam2024dramscope}.\footnote{For some modules, the even/odd column assignment could not be reliably determined 
(i.e., the expected parity-dependent pattern was not consistently observed)
using the \RowCopy{}-based method of~\cite{yuksel2024functionally,nam2024dramscope}. We exclude these modules from experiments involving even/odd column parity.}

\noindent
\textbf{\dtcr{4}{Verification of} Simultaneously Activated Rows.}
Prior work~\cite{yuksel2024functionally, yuksel2024simultaneous, yuksel2025pudhammer, yukel2025simratrng, yuksel2023pulsar} demonstrates that an \APA{} sequence can simultaneously activate 2, 4, 8, 16, and 32 DRAM rows, and a subsequent \WRITE{} command overwrites these rows with the written data.\footnote{For some modules, \SiMRA{}  is unreliable at certain activation row counts (i.e., some intended rows are not activated). We exclude such modules from the results of the affected row counts.}
We follow this methodology to verify which rows are activated during \SiMRA{}. First, we initialize an entire subarray with a predefined pattern. Second, we issue an \APA{} sequence to activate multiple target rows, immediately followed by a \WRITE{} command with a distinct pattern. After issuing a \PRE, we read back each row in the subarray. Rows that were activated during \APA{} will contain the written pattern, allowing us to identify which rows participated in \SiMRA{}. 
To ensure non-activated rows remain unchanged, we extend the delay between the \WRITE{} command following \APA{} and the subsequent \PRE{} command beyond the nominal tWR timing. We test delays of tWR, tWR+50ns, tWR+100ns, and tWR+200ns to rigorously verify that non-activated rows are not overwritten.


\subsection{Overview of Experiments}
\subsubsection{Study Scope}
We test whether a key expectation of PuD computations holds on real DRAM chips: that each column's output depends solely on its own operand data. We investigate two sources of interference from non-operand data by addressing the following research questions\ieycr{4}{, RQ1 and RQ2.}

\noindent\textbf{RQ1: How does data stored in non-activated rows affect \SiMRA{} outputs?} (\S\ref{sec:row:ide} and \S\ref{sec:row:adj})

\noindent\textbf{RQ2: How does data \dtcr{4}{stored in other columns that concurrently perform computations under the same \SiMRA{} operation affect \SiMRA{} outputs?}} (\S\ref{sec:col})

\noindent
\dtcr{5}{We provide hypothetical explanations for the phenomena observed on real DRAM chips in \S\ref{sec:hyp}.}



\subsubsection{Terminology and Metric}\label{sec:met:ovr:ter}
\begin{figure}[htbp]
  \centering
  \includegraphics[width=1.0\columnwidth]{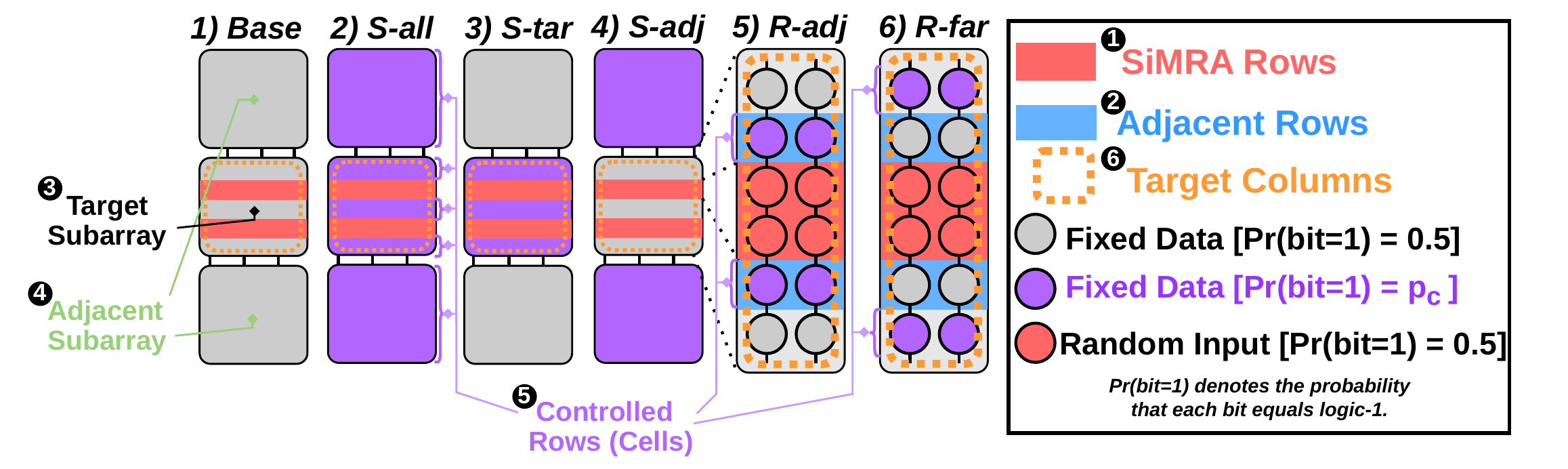}
  \caption{\dtcr{5}{Experimental setup for evaluating interference from non-activated rows.}}
  \label{fig:row-ide}
\end{figure}

\ieycr{4}{Figure~\ref{fig:row-ide} illustrates the six key terms we use in \SiMRA{} experiments.} 
\textbf{\textit{SiMRA rows}} (\blackcircled{1})
refers to a set of rows that are simultaneously activated during a \SiMRA{} operation; these rows \iey{contain} the input \iey{operand}s on which the operation is performed. 
\textbf{\textit{Adjacent rows}} (\blackcircled{2}) are the rows that are \dtcr{4}{\emph{not}}
\jwang{activated during \SiMRA{} \iey{and} physically adjacent to} the SiMRA rows.
\textbf{\textit{Target subarray}} (\blackcircled{3}) \iey{refers to} the subarray that contains the \SiMRA{} rows. 
\textbf{\textit{Adjacent subarrays}} (\blackcircled{4}) are physically adjacent \iey{subarrays of the target subarray, one} above and \iey{one} below the target subarray. 
We \iey{define \ieycr{4}{the}
term} \textbf{\textit{controlled rows}} (or \textbf{\textit{controlled cells}}) (\blackcircled{5}) to \ieycr{4}{refer to}
rows \dtcr{4}{(}or cells\dtcr{4}{)} whose data pattern is intentionally configured to observe how the \SiMRA{} output changes in response to different data patterns.
These data patterns include fixed patterns (all zeros, all ones, and random patterns) and patterns that depend on the inputs to the \SiMRA{} rows.
For random patterns, we define \(\bm{p}_{c}\) as the probability that each bit in controlled cells independently takes logic-1.
\textbf{\textit{Target columns}} (\blackcircled{6}) refer to a specified subset of columns over which we evaluate the \SiMRA{} output. \dtcr{4}{By default, all columns in the subarray serve as target columns.} For column-wise interference experiments (shown in Figure~\ref{fig:row-col} and \S\ref{sec:col}), 
\dtcr{4}{we randomly select \(1/8\)\dtcr{5}{th} of all columns in the subarray as target columns. The \SiMRA{}-row cells in the remaining \(7/8\)\dtcr{5}{th} columns serve as controlled cells, whose data patterns we vary to} \iey{analyze their impact on \SiMRA{} outputs of the target columns.}

\dtcr{4}{We define \(\bm{p}_{o1}\) as the fraction of output bits that equal logic-1 when \SiMRA{} is executed with random inputs over the target columns. For example, if target columns cover all \(64\mathrm{K}\) columns in a bank and we collect 128 samples, \(p_{o1}=0.60\) means that 60\% of the \(64\mathrm{K}\times 128\) output bits are logic-1.
We use \(p_{o1}\) to quantify \ieycr{4}{the effect of} PuDGhost \ieycr{4}{on the \SiMRA{} operation} \ieycr{4}{in three key steps}. \ieycr{4}{First, a}s a baseline, we set all data other than the operand data in the target columns to random patterns (i.e., \(p_c = 0.5\)). \ieycr{4}{Second, w}e vary the data pattern of specific controlled rows or cells and measure the resulting change in \(p_{o1}\) relative to this baseline. A change in \(p_{o1}\) indicates that the controlled non-operand data affects the computation output, and the direction and magnitude of the shift quantify the interference.}
\dtcr{4}{\ieycr{4}{Third, w}e report the ratio of \(p_{o1}\) under each condition to the baseline \(p_{o1}\), which we call \textbf{\textit{Norm.}}~\(\bm{p}_{o1}\). A Norm.~\(p_{o1}\) closer to 1.0 indicates less interference from the controlled cells, while values below (above) 1.0 indicate that the controlled cells bias the \SiMRA{} output toward logic-0 (logic-1).}
\section{Interference from Non-Activated Rows}\label{sec:row:ide}
In this section, we examine how non-activated rows in the \emph{target} and \emph{adjacent} subarrays, which share bitlines with the \SiMRA{} rows, affect \SiMRA{} outputs.

\subsection{Experimental Methodology}
\noindent\textbf{Metric.}
\dtcr{4}{We designate several rows in the target subarray and adjacent subarrays as controlled rows. For each experimental condition, we collect 128 samples with random inputs to the \SiMRA{} rows while keeping the data in the controlled rows fixed across samples. In this section, \ieycr{4}{we use all the columns in a subarray (i.e., 65536 columns) as target columns to perform \SiMRA{} operations. Thus, we generate a total of 65536$\times$128 \SiMRA{} output bits}}
\dtcr{4}{per condition. We report Norm.~\(p_{o1}\) as defined in \S\ref{sec:met:ovr:ter} to quantify the interference from the controlled rows.}

\noindent\textbf{Conditions.}
\ieycr{4}{Figure~\ref{fig:row-ide} illustrates six different conditions that we define to}
progressively \ieycr{4}{understand} which non-activated rows affect the \SiMRA{} output in the scope of three consecutive subarrays, the target subarray (\blackcircled{3}) and its adjacent subarrays (\blackcircled{4}).
\textbf{\textit{1) Base}} is the baseline configuration.
\textbf{\textit{2) S-all}} designates \dtcr{4}{all} rows except the \SiMRA{} rows in both the target (\blackcircled{3}) and adjacent subarrays (\blackcircled{4}) as controlled rows.
\textbf{\textit{3) S-tar}} designates all rows \dtcr{4}{except the \SiMRA{} rows} in the target subarray as controlled rows, while \textbf{\textit{4) S-adj}} designates \dtcr{4}{all} rows in the adjacent subarrays.
\textbf{\textit{5) R-adj}} designates \dtcr{4}{all} adjacent rows (\blackcircled{2})
as controlled rows, while \textbf{\textit{6) R-far}} designates \dtcr{4}{all} rows in the target subarray \dtcr{4}{except the \SiMRA{} rows (\blackcircled{1}) and the adjacent rows (\blackcircled{2})}.


\noindent\textbf{Experimental Control.}
Across all conditions, the 128 sets of random input data applied to the \SiMRA{} rows are \emph{identical}, and the fixed data in all \emph{non-controlled} rows are also kept identical. 
This design enables us to attribute observed differences in \({p}_{o1}\) to the data patterns in the controlled rows.

\noindent\textbf{Experimental Protocol.}
\dtcr{4}{For each of the 128 random-input samples per condition, \dtcr{5}{we execute 
four key} steps:}
(i) write the specified data patterns to all controlled rows in the target and adjacent subarrays;
(ii) write random input data to the \SiMRA{} rows;
(iii) execute \SiMRA{}; and
(iv) read out the \SiMRA{} rows to record the outputs.
\dtcr{5}{Controlled rows} retain the same data pattern across all 128 random-input samples, while random inputs to the \SiMRA{} rows are independently generated for each sample.
We ensure that the time per sample is well within the DRAM refresh window to eliminate the influence of retention-time failures.

\noindent\textbf{Number of Instances Tested.}
To keep the total testing time reasonable, for each DRAM module, we select three subarrays per bank (one from the upper, one from the middle, and one from the lower region of the bank).
Within each selected subarray, we randomly choose a group of \SiMRA{} rows for each activation count: 2-, 4-, 8-, 16-, and 32-row activation.

\noindent\textbf{Temperature.}
Unless stated otherwise, all experiments are conducted at \(50^{\circ}\mathrm{C}\).

\subsection{COTS DRAM Chip Characterization}
\begin{figure}[h]
  \centering
  \begin{subfigure}{\linewidth}
    \includegraphics[width=\linewidth]{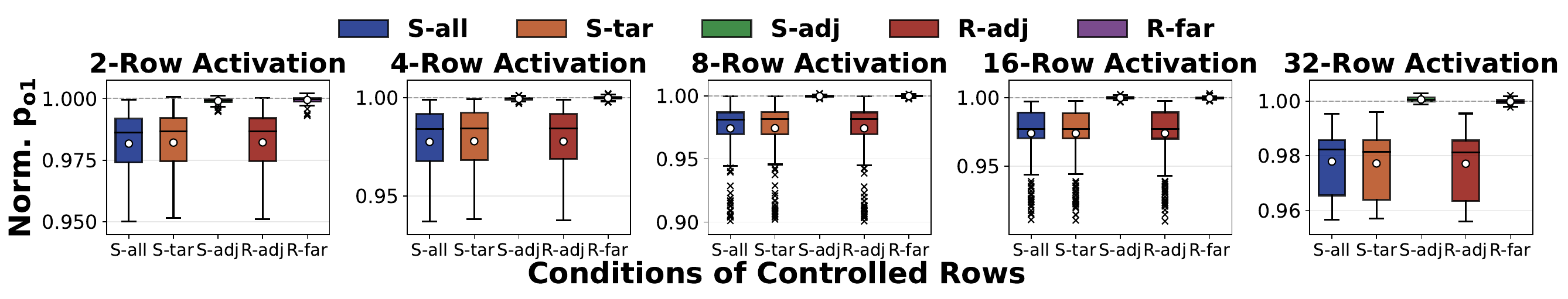}
    \caption{All controlled rows store an all-zero pattern.}
    \label{fig:row-type:zero}
  \end{subfigure}
  \begin{subfigure}{\linewidth}
    \includegraphics[width=\linewidth]{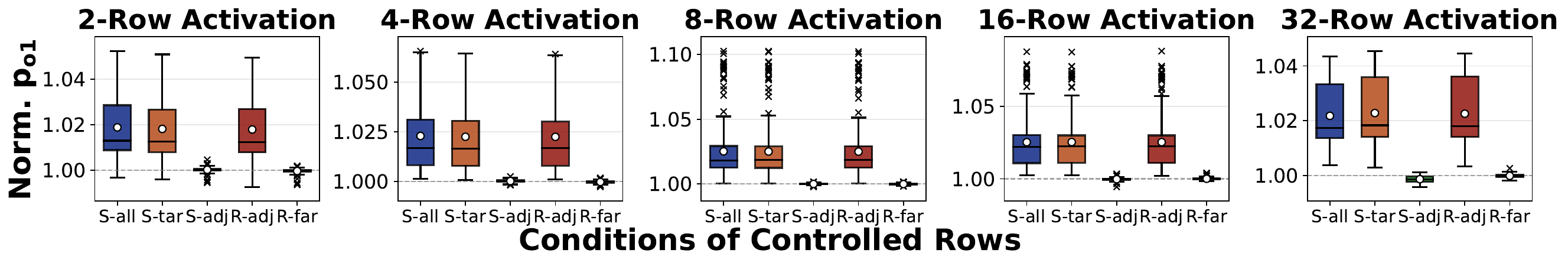}
    \caption{All controlled rows store an all-one pattern.}
    \label{fig:row-type:one}
  \end{subfigure}
  \caption{Row-type characterization under controlled patterns.}
  \label{fig:row-type}
\end{figure}

Figure \ref{fig:row-type} shows how non-activated rows affect \SiMRA{} outputs under six conditions.\footnote{\dtcr{5}{Throughout \S\ref{sec:row:ide}, \S\ref{sec:row:adj}, and \S\ref{sec:col}, box-and-whisker plots} show box boundaries representing the first and third quartiles (Q1 and Q3), and circle markers indicating the mean values.
\dtcr{5}{Each box shows the distribution across all tested instances (one \SiMRA{} row group per subarray, across 12 modules $\times$ 16 banks $\times$ 3 subarrays).}}
\dtcr{4}{The y-axis shows the distribution of Norm.~\({p}_{o1}\) across all tested modules, banks, and subarrays and the x-axis shows each tested condition.}
\dtcr{4}{\ieycr{4}{Figure~\ref{fig:row-type}a shows} 
when all controlled rows store all-zeros (\(p_c = 0.0\)), and \ieycr{4}{Figure~\ref{fig:row-type}b shows} when all controlled rows store all-ones (\(p_c = 1.0\))}. \ieycr{4}{Each subplot represents a different number of \SiMRA{} rows (i.e., 2-, 4-, 8-, 16-, and 32-row activation).} 

\begin{observation}
\dtcr{4}{Non-activated adjacent rows bias \SiMRA{} outputs toward logic-0 (logic-1) when the adjacent rows store logic-0 (logic-1).}
\end{observation}
In Figure~\ref{fig:row-type}a, we observe that S-all, S-tar, and R-adj produce Norm.~\({p}_{o1}\) of approximately 0.97--0.98 on average across all five activation counts, reaching as low as 0.90. This indicates that adjacent-row data set to all-zeros biases \SiMRA{} outputs for random inputs toward logic-0 by \dtcr{5}{an average of} 2--3\% and up to 10\%.
Similarly, in Figure~\ref{fig:row-type}b, S-all, S-tar, and R-adj produce Norm.~\({p}_{o1}\) of 1.02--1.03 on average across all five activation counts, reaching as high as 1.10. This indicates that adjacent-row data set to all-ones biases \SiMRA{} outputs for random inputs toward logic-1 by \dtcr{5}{an average of} 2--3\% and up to 10\%.

\begin{observation}
Interference from non-activated rows is highly localized to physically adjacent rows.
\end{observation}

The R-adj condition shows similar Norm.~\({p}_{o1}\) values to S-all and S-tar, with means of approximately 0.97--0.98 in Figure~\ref{fig:row-type}a and 1.02--1.03 in Figure~\ref{fig:row-type}b, demonstrating that adjacent rows account for essentially all observed interference. In contrast, conditions without adjacent-row control (S-adj and R-far) show at most 0.7\% deviation from the baseline in both Figure~\ref{fig:row-type}a and Figure~\ref{fig:row-type}b, indicating that non-adjacent rows have a small impact on \SiMRA{} outputs.

\begin{observation}
Interference from \dtcr{4}{non-activated} adjacent rows can strengthen as the number of \SiMRA{} rows increases.
\end{observation}
We observe that the magnitude of bias can increase with more \SiMRA{} rows. For example, in the R-adj condition, Norm.~\({p}_{o1}\) reaches as low as 0.95 in Figure~\ref{fig:row-type}a and as high as 1.05 in Figure~\ref{fig:row-type}b for 2-row activation, while for 8-row activation, it reaches as low as 0.90 in Figure~\ref{fig:row-type}a and as high as 1.10 in Figure~\ref{fig:row-type}b.




\section{Characterization of Interference from Adjacent Rows}\label{sec:row:adj}
Building on the result of \S\ref{sec:row:ide}, this section provides a detailed characterization of adjacent-row interference from several perspectives (e.g., data patterns, spatial locality, and temperature sensitivity). We reuse the experimental control, the number of instances, and the default temperature setup from \S\ref{sec:row:ide} and describe only the methodology specific to \dtcr{5}{the} experiment.

\subsection{Experimental Methodology}

\noindent\textbf{Metric.}
As in the R-adj setup in Figure~\ref{fig:row-ide}, we designate all adjacent rows (\blackcircled{2}) as controlled rows (\blackcircled{5}) and use \dtcr{5}{normalized} (Norm.) \({p}_{o1}\) to quantitatively evaluate their impact on the \SiMRA{} output (the common baseline condition fills all adjacent rows with independent random data with \(p_c=0.5\)). 
For each experimental condition, we collect 128 samples with random inputs to the \SiMRA{} rows (\blackcircled{1}). Except for the experiment shown in Figure~\ref{fig:row_pair}, the data in the controlled rows (\blackcircled{5}) are kept fixed across all samples.
Unless otherwise noted, all columns are designated as target columns (\blackcircled{6}).

\noindent\textbf{Experimental Protocol.}
For each of the 128 random-input samples, we execute the following four steps: (i) write the specified data pattern to all adjacent rows; (ii) write random input data to the \SiMRA{} rows; (iii) execute \SiMRA{}; and (iv) read out the \SiMRA{} rows to record the outputs.


\subsection{COTS DRAM Chip Characterization Results}
\begin{figure}[h]
    \centering
    \includegraphics[width=\columnwidth]{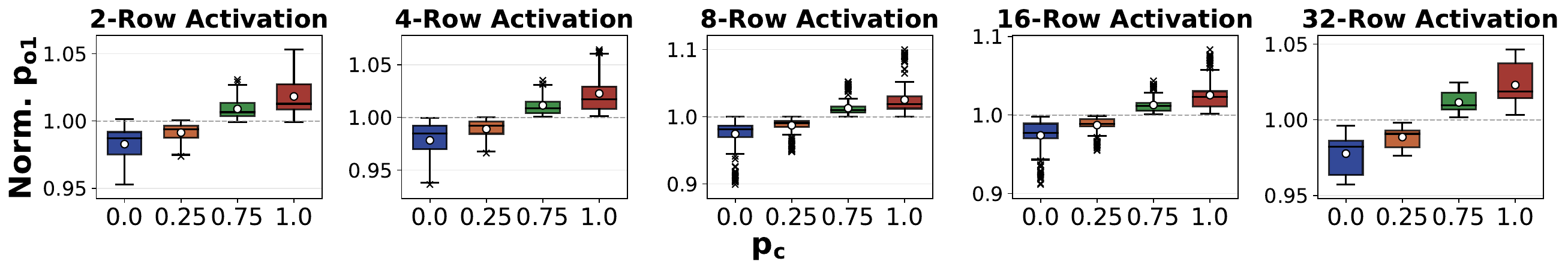}
    \caption{Sensitivity to fraction of logic-1 in adjacent rows.}
    \label{fig:row}
\end{figure}

\noindent\textbf{\textit{Sensitivity to Fraction of Logic-1.}}
To examine how the fraction of logic-1 stored in adjacent rows affects the \SiMRA{} output, we sweep \(p_c \in \{0.0,\, 0.25,\, 0.75,\, 1.0\}\).
Note that \(p_c=0.5\) corresponds to the baseline condition.
Figure~\ref{fig:row} shows Norm.~\({p}_{o1}\) distribution (y-axis) for each \(p_c\) value of the adjacent rows (x-axis), across different numbers of \SiMRA{} rows (each subplot).

\begin{observation}\label{obs:adj:log}
Data stored in non-activated adjacent rows more strongly biases \SiMRA{} outputs toward logic-1 (logic-0) as the fraction of logic-1 (logic-0) in the adjacent rows increases.
\end{observation}

As \(p_c\) increases, Norm.~\({p}_{o1}\) increases monotonically, demonstrating a clear positive correlation between the fraction of logic-1 in adjacent rows and the bias in \SiMRA{} outputs. For example, with 32-row activation, the mean Norm.~\({p}_{o1}\) rises monotonically from 0.98 (\(p_c = 0.0\)) to 1.02 (\(p_c = 1.0\)).

\begin{figure}[H]
    \centering
    \includegraphics[width=\columnwidth]{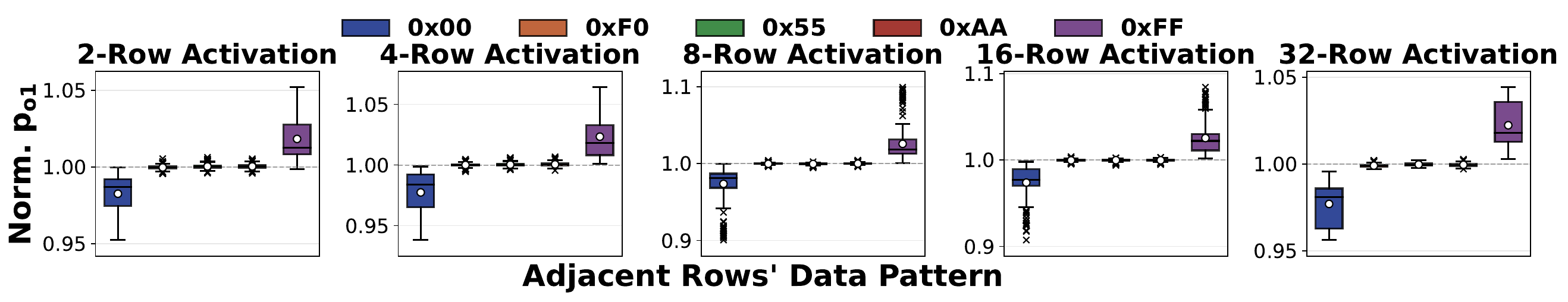}
    \caption{Sensitivity to the data patterns in the adjacent rows.}
    \label{fig:row_data}
\end{figure}

\noindent\textbf{\textit{Structured Data Patterns.}}
To study the impact of structured (i.e., non-random) data patterns in adjacent rows, we vary the byte pattern among \(\{ \texttt{0x00}, \texttt{0xF0}, \texttt{0x55}, \texttt{0xAA}, \texttt{0xFF} \}\).
Figure~\ref{fig:row_data} shows Norm.~\({p}_{o1}\) distribution (y-axis) for various data patterns written to the adjacent rows (x-axis), across different numbers of \SiMRA{} rows (each subplot).

\begin{observation}
Non-activated adjacent rows bias \SiMRA{} outputs based on the fraction of logic-1 they store rather than on the specific structured data pattern.
\end{observation}

Data patterns with the same fraction of logic-1 in adjacent rows produce similar Norm.~\({p}_{o1}\) regardless of the structured pattern used. 0xF0, 0x55, and 0xAA all have a fraction of logic-1 of 0.5 and produce Norm.~\({p}_{o1}\) of approximately 1.0.

\begin{figure}[h]
    \centering
    \includegraphics[width=0.9\columnwidth]{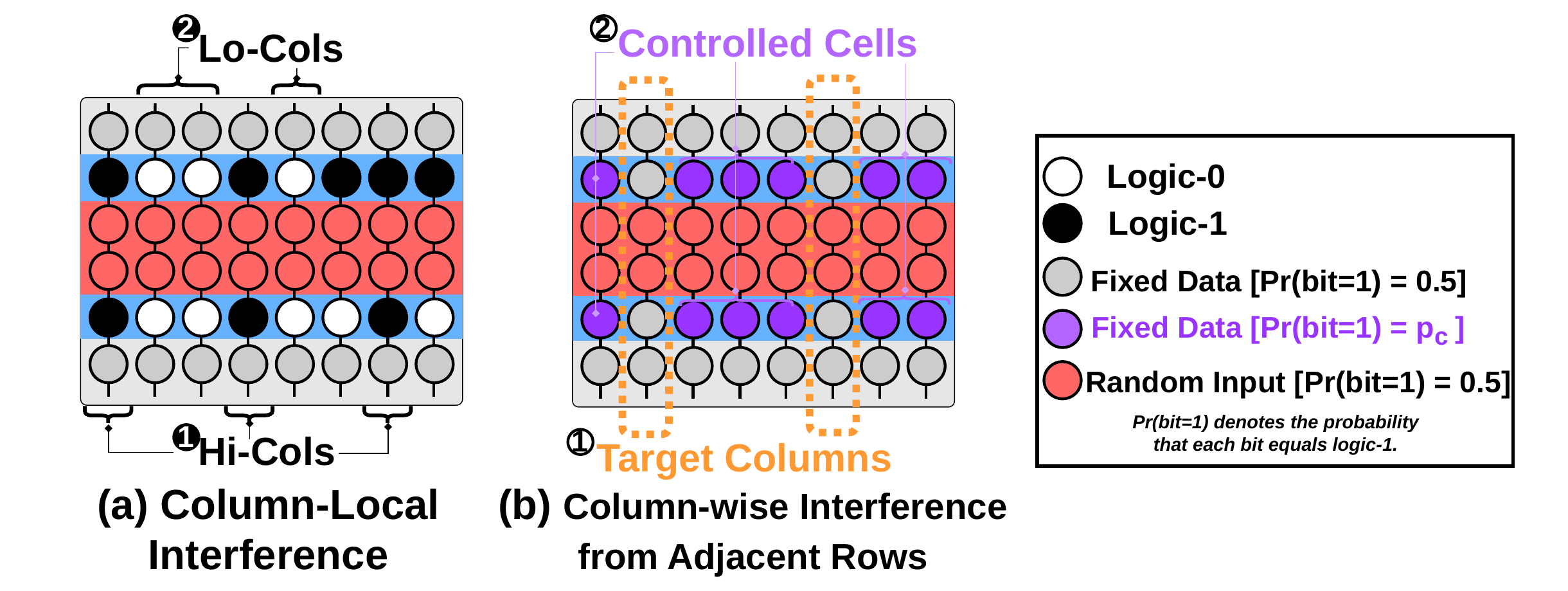}
    \caption{(a) Column-local interference. (b) Adjacent-row interference across columns.}
    \label{fig:row-col}
\end{figure}

\noindent\textbf{\textit{Column-Local Interference.}}
We test whether interference from adjacent rows is uniform across all columns or depends on each column's own adjacent-row cells.
We set data in adjacent rows to random (\(p_c = 0.5\)) and partition columns into two sets based on the fraction of logic-1 in each column's adjacent-row cells (see Figure~\ref{fig:row-col}a):
\blackcircled{1}~\emph{Hi-Cols} (columns where more than half of the adjacent-row cells store logic-1) and \blackcircled{2}~\emph{Lo-Cols} (columns where fewer than half of the adjacent-row cells store logic-1). \dtcr{5}{Columns where the adjacent-row cells store the same number of logic-0 and logic-1 are not classified into either \blackcircled{1} or \blackcircled{2}.}
We measure Norm.~\({p}_{o1}\) separately for Hi-Cols and Lo-Cols as target columns.

\begin{figure}[!h]
    \centering
    \includegraphics[width=\columnwidth]{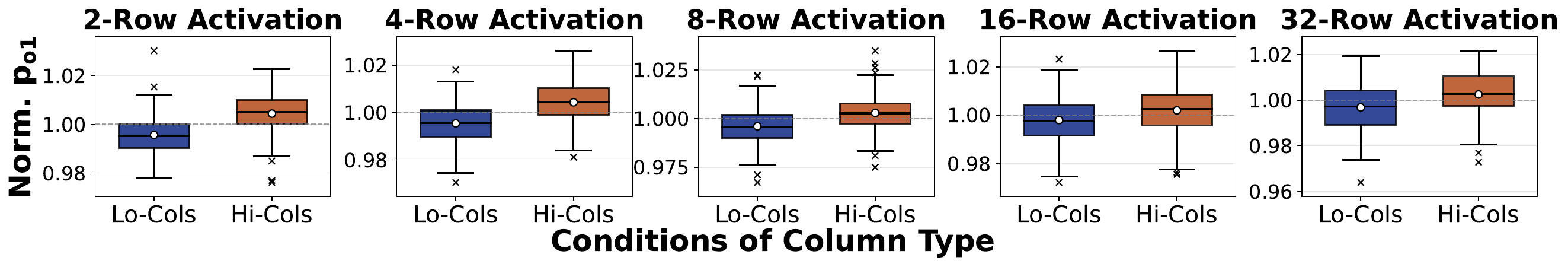}
    \caption{Column-local \dtcr{5}{interference}.}
    \label{fig:row_colspec}
\end{figure}

\noindent
Figure~\ref{fig:row_colspec} shows the Norm.~\({p}_{o1}\) distribution (y-axis) separately for Hi-Cols and Lo-Cols (x-axis) across different numbers of \SiMRA{} rows (each subplot).

\begin{observation}
Interference from non-activated adjacent rows is not uniform across all columns: each column's \SiMRA{} output is affected by its own adjacent-row cells.
\end{observation}

Hi-Cols and Lo-Cols show opposite trends: Hi-Cols produce a mean Norm.~\({p}_{o1}\) above 1.0 while Lo-Cols produce below 1.0. For example, with 2-row activation, the mean Norm.~\({p}_{o1}\) is 1.004 for Hi-Cols and 0.996 for Lo-Cols. This means that columns whose own adjacent-row cells contain more logic-0 (logic-1) are biased toward logic-0 (logic-1), confirming that adjacent-row interference is not uniform across all columns.

\begin{figure}[h]
    \centering
    \includegraphics[width=\columnwidth]{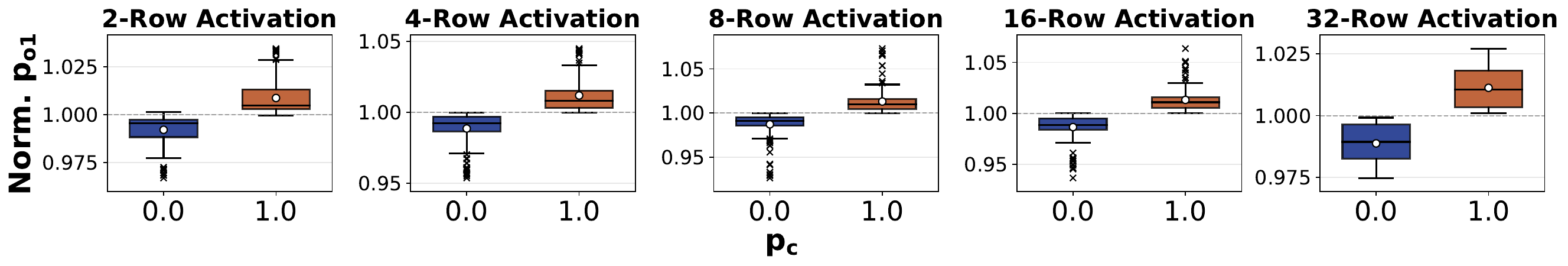}
    \caption{Adjacent-row interference across columns.}
    \label{fig:col_adjrow_fast}
\end{figure}

\noindent
\noindent\textbf{\textit{Adjacent-Row Interference Across Columns.}}
To evaluate whether a given column's \SiMRA{} output is affected by data stored in other columns' adjacent-row cells, we randomly select \(1/8\)\dtcr{5}{th} of all columns as target columns (\circledt{white}{1}) (see Figure~\ref{fig:row-col}b).
For the remaining \(7/8\)\dtcr{5}{th} columns, we designate their adjacent-row cells as controlled cells (\circledt{white}{2}) and set them to all-zeros (\(p_c = 0.0\)) or all-ones (\(p_c = 1.0\)).
Figure~\ref{fig:col_adjrow_fast} shows the Norm.~\({p}_{o1}\) distribution of the target columns (y-axis) for different \(p_c\) values of the controlled cells in non-target columns (x-axis), across different numbers of \SiMRA{} rows (each subplot).

\begin{observation}
A given column's \SiMRA{} output is affected by data stored in other columns' adjacent-row cells.
\end{observation}

When the controlled cells in non-target columns store logic-0 (logic-1), the mean Norm.~\({p}_{o1}\) of target columns is biased toward logic-0 (logic-1), confirming that data in other columns' adjacent-row cells affects a given column's \SiMRA{} output. For example, with 2-row activation, the mean Norm.~\({p}_{o1}\) shifts to 0.99 at \(p_c = 0.0\) and 1.01 at \(p_c = 1.0\).

\begin{figure}[h]
    \centering
    \includegraphics[width=0.93\columnwidth]{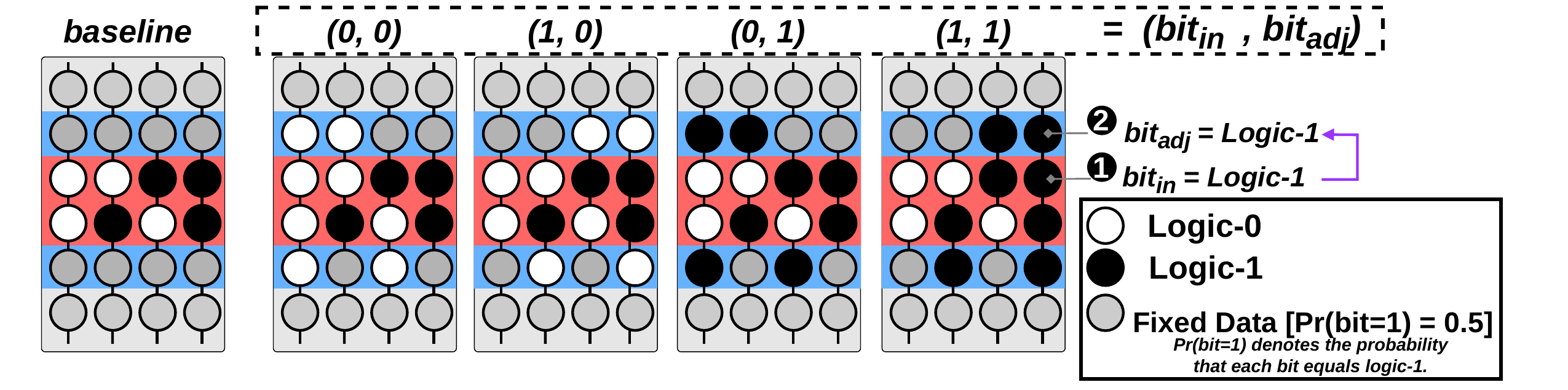}
    \caption{Input-dependent adjacent-row patterns.}
    \label{fig:row-pair}
\end{figure}

\noindent
\noindent\textbf{\textit{Input-Dependent Adjacent-Row Patterns.}}
To quantify how a \SiMRA{} input bit and its corresponding adjacent-row bit jointly affect the output, \dtcr{5}{we define \(bit_{\text{in}}\) as the value of a \SiMRA{} input bit in a given column and \(bit_{\text{adj}}\) as the value of the adjacent-row cell in the same column.}
For each random-input sample, we first initialize all adjacent rows with a random base pattern (\(p_c = 0.5\)). We then modify the adjacent-row cells based on a specified \((bit_{\text{in}}, bit_{\text{adj}})\) combination: for each column where the \SiMRA{} input equals \(bit_{\text{in}}\), we set its corresponding adjacent-row cell to \(bit_{\text{adj}}\); all other adjacent-row cells retain the base pattern.
Figure~\ref{fig:row-pair} illustrates the adjacent-row data patterns for the baseline and all four combinations in a given random-input sample. For example, for \((bit_{\text{in}}, bit_{\text{adj}}) = (1, 1)\), each column where the \SiMRA{} input is logic-1 (\blackcircled{1}) has its corresponding adjacent-row cell set to logic-1 (\blackcircled{2}), while the remaining adjacent-row cells retain the random base pattern.
Because the random inputs differ across samples, the spatial distribution of modified adjacent-row cells changes accordingly.
This experiment uses configurations where each adjacent row borders only one \SiMRA{} row, making \(bit_{\text{adj}}\) well-defined.

\begin{figure}[h]
    \centering
    \includegraphics[width=\columnwidth]{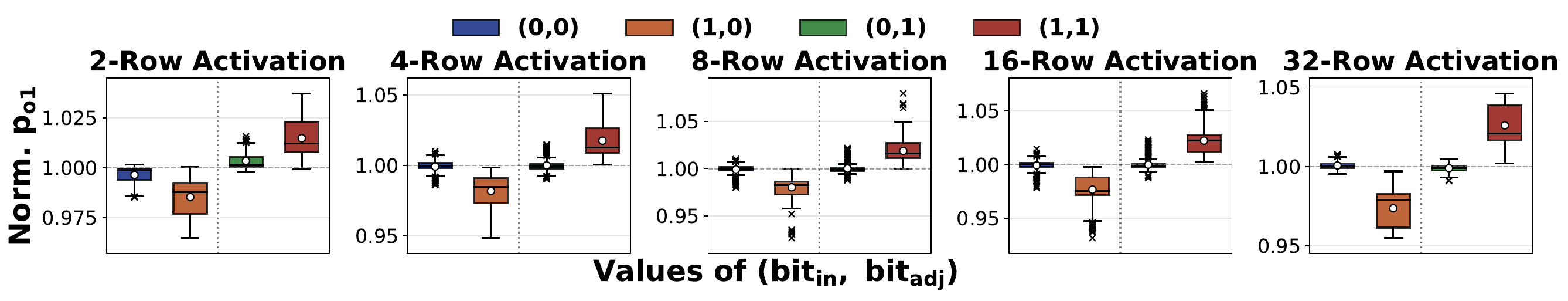}
    \caption{Input-dependent adjacent-row interference.}
    \label{fig:row_pair}
\end{figure}

\noindent
Figure~\ref{fig:row_pair} shows Norm.~\({p}_{o1}\) distribution (y-axis) for all four combinations \dtcr{5}{\((bit_{\text{in}}, bit_{\text{adj}}) \in \{(0,0),\, (1,0),\, (0,1),\, (1,1)\}\)} (x-axis), across different numbers of \SiMRA{} rows (each subplot).

\begin{observation}\label{obs:adj:pair}
Cells in simultaneously activated rows storing logic-1 are substantially more susceptible to interference from non-activated adjacent rows than cells storing logic-0.
\end{observation}

When \(bit_{\text{in}} = 0\), the mean Norm.~\({p}_{o1}\) remains approximately 1.0 regardless of \(bit_{\text{adj}}\) across all tested numbers of \SiMRA{} rows, indicating negligible interference. In contrast, when \(bit_{\text{in}} = 1\), the \SiMRA{} output is sensitive to the adjacent-row cell: \(bit_{\text{adj}} = 0\) decreases the mean Norm.~\({p}_{o1}\) to approximately 0.97--0.98, while \(bit_{\text{adj}} = 1\) increases it to 1.02--1.03.



\begin{figure}[h]
    \centering
    \includegraphics[width=\columnwidth]{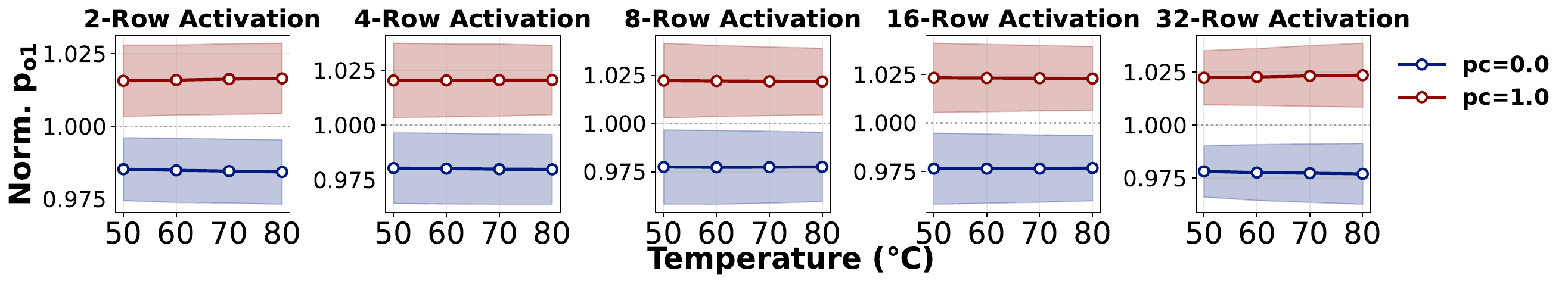}
    \caption{\dtcr{5}{Adjacent-row interference vs. temperature.}}
    \label{fig:row_temp}
\end{figure}

\noindent
\noindent\textbf{\textit{Temperature Sensitivity.}}
To study the temperature dependence of interference from adjacent rows, we set the temperature to \(50^{\circ}\mathrm{C}\), \(60^{\circ}\mathrm{C}\), \(70^{\circ}\mathrm{C}\), and \(80^{\circ}\mathrm{C}\).
At each temperature, we set all adjacent rows to all-zeros (\(p_c = 0.0\)) or all-ones (\(p_c = 1.0\)).
Figure~\ref{fig:row_temp} shows the mean Norm.~\({p}_{o1}\) (y-axis) across temperatures ranging from \(50^{\circ}\mathrm{C}\) to \(80^{\circ}\mathrm{C}\) (x-axis) when all adjacent rows are set to all-zeros (\(p_c = 0.0\)) or all-ones (\(p_c = 1.0\)), across different numbers of \SiMRA{} rows (each subplot).\footnotemark

\begin{observation}
\dtcr{5}{Temperature} has a small effect on interference from non-activated adjacent rows.
\end{observation}

We observe no significant temperature dependence in adjacent-row interference. For example, with 8-row activation under \(p_c = 1.0\), the mean Norm.~\({p}_{o1}\) remains at approximately 1.02 across all tested temperatures (\(50^{\circ}\mathrm{C}\)--\(80^{\circ}\mathrm{C}\)).

\footnotetext{\dtcr{5}{Throughout \S\ref{sec:row:adj} and \S\ref{sec:col},} the line plots show mean values with shaded bands representing the interquartile range (IQR), \dtcr{5}{computed across all tested instances (one \SiMRA{} row group per subarray, across 12 modules $\times$ 16 banks $\times$ 3 subarrays)}.}
\section{Interference from Columns that Concurrently Perform Computations}\label{sec:col}
In this section, we investigate whether data stored in columns that concurrently perform computations under the same \SiMRA{} operation affects the \SiMRA{} outputs.

\subsection{Experimental Methodology}
We reuse the experimental protocol, the experimental control, the number of instances, and the temperature setup from \S\ref{sec:row:adj} and describe only the methodology specific to \S\ref{sec:col}.

\begin{figure}[h]
    \centering
    \includegraphics[width=\columnwidth]{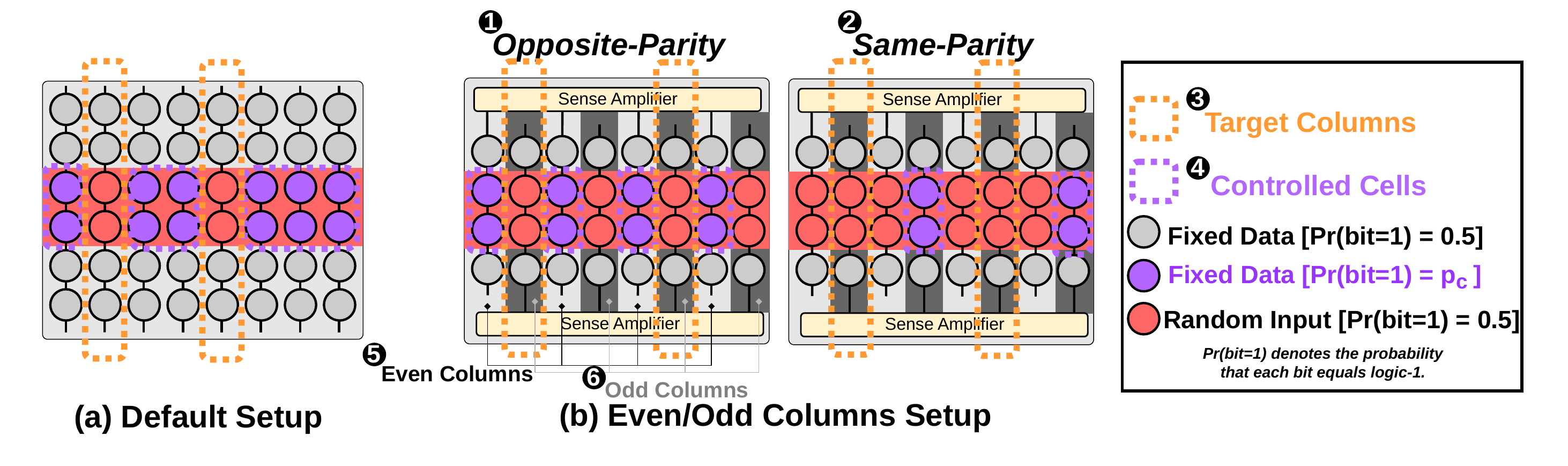}
    \caption{Experimental setup for evaluating interference from concurrently computing columns\dtcr{5}{:} (a) Default setup\dtcr{5}{,} (b) Even/odd column setup.}
    \label{fig:col-inp}
\end{figure}

\noindent\textbf{Metric.}
We select \(1/8\)\dtcr{5}{th} of all columns as target columns and designate the \SiMRA{}-row cells in the remaining \(7/8\)\dtcr{5}{th} columns as controlled cells.
By default, target columns are randomly selected from all columns regardless of even/odd column parity, as illustrated in Figure~\ref{fig:col-inp}a.
We measure Norm.~\({p}_{o1}\) over the target columns using 128 random-input samples.
The baseline condition sets each controlled cell independently to logic-1 with probability \(p_c=0.5\).

\subsection{COTS DRAM Chip Characterization Results}
\noindent\textbf{\textit{Fraction of Logic-1.}}
Using the default setup (Figure~\ref{fig:col-inp}a), we sweep \(p_c \in \{0.0,\, 0.25,\, 0.75,\, 1.0\}\) to examine how the fraction of logic-1 in the controlled cells affects the \SiMRA{} outputs of target columns. Note that \(p_c=0.5\) corresponds to the baseline.

\begin{figure}[H]
    \centering
    \includegraphics[width=\columnwidth]{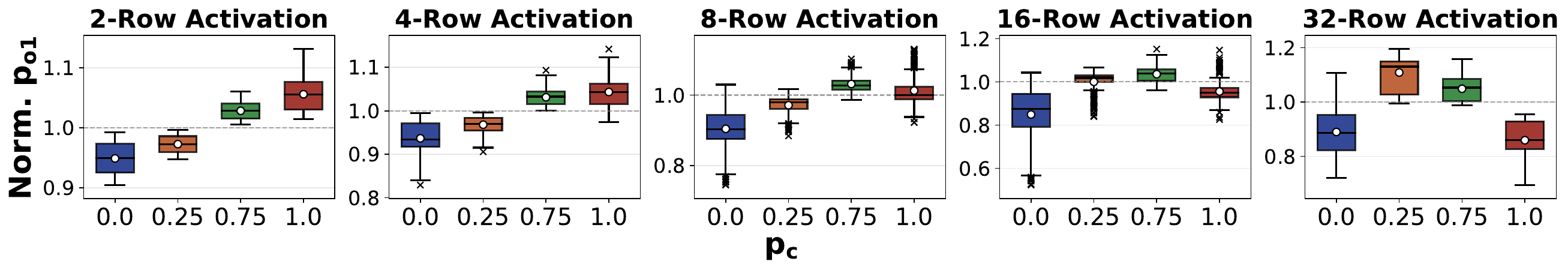}
    \caption{Sensitivity to the fraction of logic-1 in concurrently computing columns' inputs.}
    \label{fig:col_fast}
\end{figure}

Figure~\ref{fig:col_fast} shows the Norm.~\({p}_{o1}\) distribution of the target columns (y-axis) across all tested modules, banks, and subarrays, for each \(p_c\) value of the controlled cells (x-axis), across different numbers of \SiMRA{} rows (each subplot).

\begin{observation}\label{obs:col:log}
\SiMRA{} output is affected by data stored in other columns that concurrently perform computations. 
\dtcr{5}{For 2--4 simultaneously activated rows, these columns' inputs bias \SiMRA{} outputs toward logic-1 (logic-0) as their fraction of logic-1 increases (decreases). For 8+ rows, the bias direction reverses at high fractions of logic-1, where \SiMRA{} outputs are biased toward logic-0.}
\end{observation}

The observed trend of the interference from columns that concurrently perform computations varies with the number of simultaneously activated rows.
For small row counts (2--4 rows), Norm.~\({p}_{o1}\) increases monotonically with the fraction of logic-1 in the inputs of columns that concurrently perform computations. For example, with 4-row activation, the mean Norm.~\({p}_{o1}\) increases from 0.94 (\(p_c = 0.0\)) to 1.04 (\(p_c = 1.0\)).
For larger row counts (8+ rows), the relationship becomes non-monotonic: Norm.~\({p}_{o1}\) peaks at intermediate \(p_c\) values and decreases toward both extremes. For example, with 32-row activation, the mean Norm.~\({p}_{o1}\) is 0.89, 1.11, 1.05, and 0.86 at \(p_c = 0.0\), \(0.25\), \(0.75\), and \(1.0\), respectively, with the maximum deviation occurring at \(p_c = 0.25\).

\begin{observation}
Interference from columns that concurrently perform computations strengthens as the number of simultaneously activated rows increases.
\end{observation}

The magnitude of interference grows substantially with more \SiMRA{} rows. For 2-row activation, the mean Norm.~\({p}_{o1}\) ranges from approximately 0.95 to 1.06. For 32-row activation, it ranges from approximately 0.86 to 1.11. \dtcr{5}{The maximum deviation from the baseline reaches up to 48\%, observed at 16-row activation with \(p_c = 0.0\).}

\begin{observation}
Interference from columns that concurrently perform computations is stronger than interference from non-activated adjacent rows.
\end{observation}

\dtcr{5}{The interference observed in Figure~\ref{fig:col_fast} is} significantly larger than the interference from non-activated rows observed in Figure~\ref{fig:row}, where the mean Norm.~\({p}_{o1}\) stays within the range of approximately 0.97--1.03.

\noindent\textbf{\textit{Even/Odd Columns.}} 
To understand the spatial characteristics of inter-column interference, we distinguish between even and odd columns. \dtcr{5}{We refer to whether a column is even or odd as its \emph{parity}; two columns have \emph{opposite parity} if one is even and the other is odd, and \emph{same parity} if both are even or both are odd.}
We select target columns from \dtcr{5}{only even columns or only odd columns}, randomly choosing \(1/8\)\dtcr{5}{th} of all columns (i.e., \(1/4\)\dtcr{5}{th} of the columns of the chosen parity) as target columns, and define two configurations, illustrated in Figure~\ref{fig:col-inp}b.
In \textbf{\textit{Opposite-Parity}} (\blackcircled{1}), the controlled cells are the \SiMRA{}-row cells in all columns of the opposite parity (e.g., in Figure~\ref{fig:col-inp}b, target columns (\blackcircled{3}) are selected from odd columns (\blackcircled{6}), so the \SiMRA{}-row cells in all even columns (\blackcircled{5}) serve as controlled cells (\blackcircled{4})).
In \textbf{\textit{Same-Parity}} (\blackcircled{2}), the controlled cells are the \SiMRA{}-row cells in all non-target columns of the same parity (e.g., in Figure~\ref{fig:col-inp}b, target columns (\blackcircled{3}) are selected from odd columns (\blackcircled{6}), so the \SiMRA{}-row cells in all non-target odd columns serve as controlled cells (\blackcircled{4})).
We sweep \(p_c \in \{0.0,\, 0.25,\, 0.75,\, 1.0\}\) for both configurations.

\begin{figure}[h]
  \centering
  \begin{subfigure}{\linewidth}
    \includegraphics[width=\linewidth]{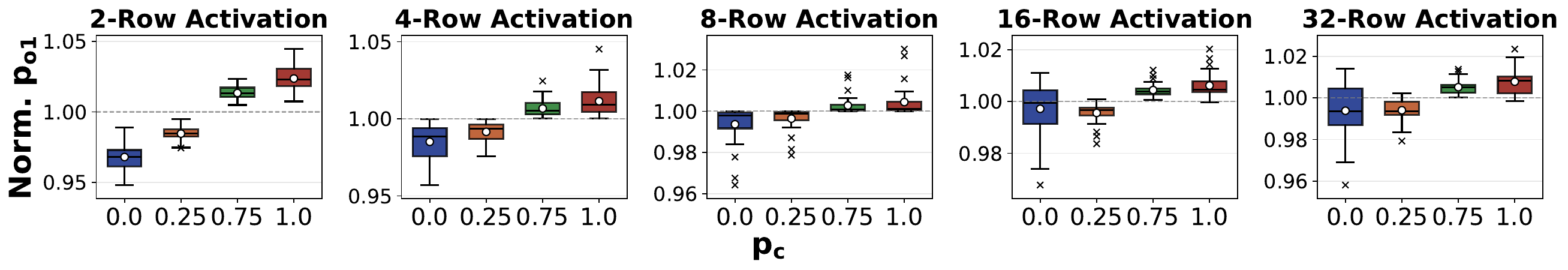}
    \caption{Opposite-Parity configuration.}
    \label{fig:col_eo:opp}
  \end{subfigure}
  \begin{subfigure}{\linewidth}
    \includegraphics[width=\linewidth]{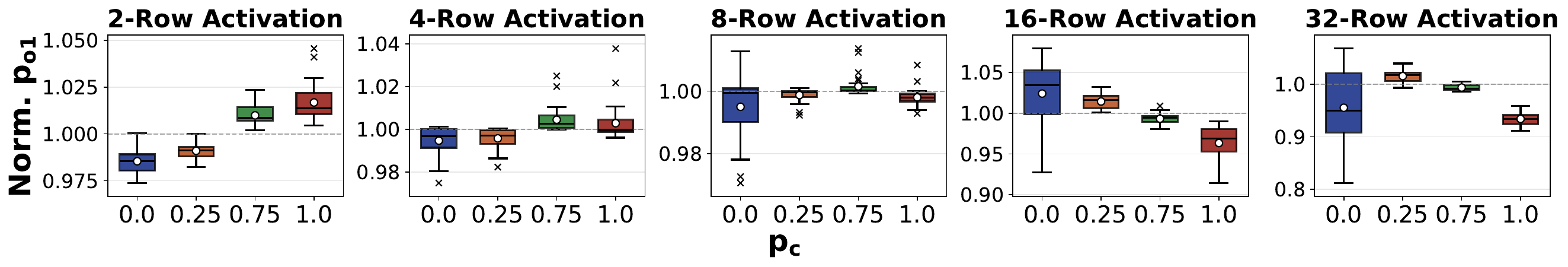}
    \caption{Same-Parity configuration.}
    \label{fig:col_eo:same}
  \end{subfigure}
  \caption{\dtcr{5}{Interference from concurrently computing columns by parity: (a) Opposite-Parity, (b) Same-Parity.}}
  \label{fig:col_eo}
\end{figure}

Figure~\ref{fig:col_eo} shows the Norm.~\({p}_{o1}\) distribution of the target columns (y-axis) for each \(p_c\) value of the controlled cells (x-axis), across different numbers of \SiMRA{} rows (each subplot). Figure~\ref{fig:col_eo:opp} shows results for the Opposite-Parity configuration, and Figure~\ref{fig:col_eo:same} for the Same-Parity configuration.

\begin{observation}
Interference from \dtcr{5}{columns of the opposite parity} biases \SiMRA{} outputs toward logic-0 (logic-1) when their inputs store logic-0 (logic-1).
\end{observation}
In Figure~\ref{fig:col_eo:opp}, the mean Norm.~\({p}_{o1}\) of the target columns increases monotonically with \(p_c\) across all five activation counts in the Opposite-Parity configuration. For example, with 2-row activation, the mean Norm.~\({p}_{o1}\) increases from approximately 0.97 at \(p_c = 0.0\) to approximately 1.02 at \(p_c = 1.0\).

\begin{observation}
Interference from \dtcr{5}{columns of the same parity} biases \SiMRA{} outputs in the same direction as their inputs for 2--4 activated rows, but the bias direction reverses at high fractions of logic-1 for 8+ rows.
\end{observation}
\dtcr{5}{In the Same-Parity configuration (Figure~\ref{fig:col_eo:same}),} for 2-row activation, the mean Norm.~\({p}_{o1}\) increases monotonically from approximately 0.99 (\(p_c = 0.0\)) to 1.02 (\(p_c = 1.0\)). 
For larger row counts (8+ rows), the bias direction can reverse at high fractions of logic-1: for example, with 32-row activation, the mean Norm.~\({p}_{o1}\) is 0.96, 1.02, 0.99, and 0.93 at \(p_c = 0.0\), \(0.25\), \(0.75\), and \(1.0\), respectively.
\dtcr{5}{This non-monotonic trend contrasts with the Opposite-Parity configuration, which remains monotonic across all row counts.}

\begin{figure}[H]
    \centering
    \includegraphics[width=\columnwidth]{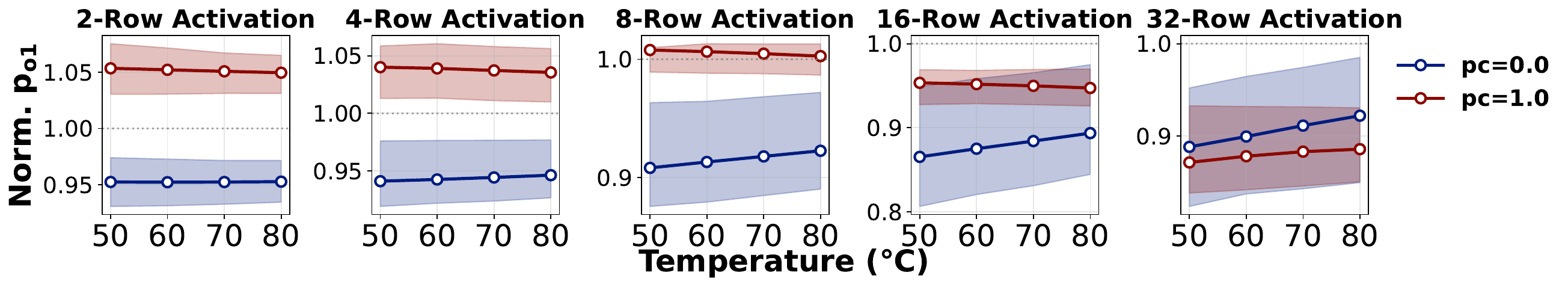}
    \caption{\dtcr{5}{Inter-column interference vs. temperature.}}
    \label{fig:col_temp}
\end{figure}

\noindent
\textbf{\textit{Temperature Sensitivity.}}
To study the temperature dependence of inter-column interference, we use the default setup (Figure~\ref{fig:col-inp}a).
At each temperature, we set the controlled cells to all-zeros (\(p_c = 0.0\)) or all-ones (\(p_c = 1.0\)).
Figure~\ref{fig:col_temp} shows the mean Norm.~\({p}_{o1}\) (y-axis) across temperatures ranging from \(50^{\circ}\mathrm{C}\) to \(80^{\circ}\mathrm{C}\) (x-axis), for \(p_c = 0.0\) and \(p_c = 1.0\) (each line), across different numbers of \SiMRA{} rows (each subplot).

\begin{observation}\label{obs:col:temp}
Interference from columns that concurrently perform computations varies monotonically with the temperature. 
\end{observation}
The interference varies monotonically with temperature. For example, with 16-row activation at \(p_c = 0.0\), the mean Norm.~\({p}_{o1}\) increases from 0.87 to 0.89 as temperature rises from \(50^{\circ}\mathrm{C}\) to \(80^{\circ}\mathrm{C}\), reducing the deviation from the baseline. In contrast, with 16-row activation at \(p_c = 1.0\), the mean Norm.~\({p}_{o1}\) decreases from 0.953 to 0.947, slightly increasing the deviation from the baseline.
\section{Hypothetical Explanation for PuDGhost}\label{sec:hyp}
We hypothesize that the interference phenomena observed in \S\ref{sec:row:ide}--\S\ref{sec:col} arise from electrical coupling during \SiMRA{} charge sharing and sensing.

\noindent\textbf{\textit{Interference from non-activated rows.}}
The directional bias observed in \S\ref{sec:row:adj} (i.e., adjacent rows storing logic-0 (logic-1) bias \SiMRA{} outputs toward logic-0 (logic-1) \dtcr{5}{(Obsv.~\ref{obs:adj:log})}) suggests electrical coupling between activated and non-activated rows, where the charge state of non-activated adjacent-row cells affects the charge sharing process.
We hypothesize that the asymmetric sensitivity observed in Figure~\ref{fig:row_pair}, where cells in \SiMRA{} rows storing logic-1 are more susceptible to interference from adjacent rows than cells storing logic-0 \dtcr{5}{(Obsv.~\ref{obs:adj:pair})}, arises because cells storing logic-1 (i.e., charged cells) have their charge drained to the bitline during \SiMRA{}, which may make them more vulnerable to interference.

\noindent\textbf{\textit{Interference from columns that concurrently perform computations.}}
We hypothesize that the monotonic trend in the Opposite-Parity configuration (Figure~\ref{fig:col_eo:opp}) arises from electrical coupling between bitlines of opposite parity.
The non-monotonic trend in the Same-Parity configuration (Figure~\ref{fig:col_eo:same}) may reflect interactions through sense-amplifier (SA) circuitry, as columns of the same parity connect to SAs on the same side of the subarray.
We hypothesize that the overall patterns observed in Figure~\ref{fig:col_fast} (Obsv.~\ref{obs:col:log}) emerge from the combination of these two effects.

We hope that our findings motivate future device-level studies to uncover the precise root causes of PuDGhost, similar to how device-level studies~\dramdevCite{} provided insight into RowHammer~\cite{kim2014flipping,kim2020revisiting,orosa2021deeperlook} and RowPress~\cite{luo2023rowpress,luo2024combinedrowhammerrowpress,luo2025revisitingreaddisturbance} after \dtcr{5}{the two phenomena were first demonstrated and analyzed in real DRAM chips}.
\section{Challenges For Reliable PuD Systems}\label{sec:rel}
Building on the detailed characterization of PuDGhost in \S\ref{sec:row:ide}, \S\ref{sec:row:adj}, and \S\ref{sec:col}, this section discusses key implications for designing future PuD systems that are robust. 

\subsection{Reliability Challenges Induced by Non-Operand Data}\label{sec:imp:err}
We demonstrate that \SiMRA{} outputs are affected by interference from data stored in non-activated adjacent rows and columns that concurrently perform computations. These results \dtcr{5}{obtained on real DRAM chips} violate the expectation of PuD computations that each column's computation should depend \emph{solely} on its own operand data.\footnotemark

Even when PuD system architects are aware that PuD computations can produce errors (e.g., due to process variation) and employ solutions such as post-manufacturing column screening to use only reliable columns, a lack of awareness of PuDGhost can still lead to unreliable systems. We demonstrate in \S\ref{sec:cas:cse} that \dtcr{5}{column screening without accounting for PuDGhost} can mislabel unreliable columns as reliable, as the non-operand data during screening may not reflect the actual non-operand data during PuD execution.

\subsection{Ineffectiveness of Disturbance Mitigations}\label{sec:imp:rhm}
PuDGhost is qualitatively different from \dtcr{5}{DRAM} read disturbance phenomena~\cite{kim2014flipping, luo2023rowpress, yuksel2025columndisturb, yuksel2025pudhammer,luo2025revisitingreaddisturbance, luo2024combinedrowhammerrowpress} in three key aspects.

\noindent\textbf{Interference source.} Read disturbance is caused by an activated aggressor row. PuDGhost is caused by data stored in 1)~\dtcr{5}{\emph{non-activated}} adjacent rows and 2)~columns that concurrently perform computations under the same \SiMRA{} operation.

\noindent\textbf{Triggering mechanism.} Read disturbance requires repeatedly activating an aggressor row (RowHammer) or keeping it open for a prolonged period (RowPress). PuDGhost requires no aggressor row activation; data stored in \dtcr{5}{\emph{non-activated}} rows affects computation results within a single \SiMRA{} operation.

\noindent\textbf{Manifestation.} Read disturbance manifests as persistent bitflips in victim DRAM cells through gradual charge injection or leakage. PuDGhost manifests as transient errors in \SiMRA{} outputs during charge sharing and sensing.

\begin{figure}[H]
    \centering
    \includegraphics[width=\columnwidth]{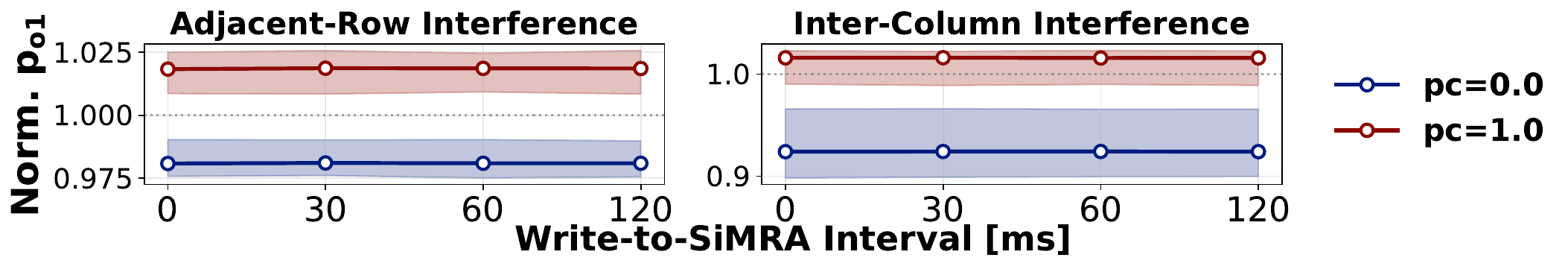}
    \caption{Time interval between writing data and \SiMRA{}.}
    \label{fig:ref}
\end{figure}

These differences render prior read disturbance mitigations~\refthrCite{} ineffective against PuDGhost.
\dtcr{5}{To experimentally verify that refreshing data before \SiMRA{} execution does \emph{not} mitigate PuDGhost, we vary the time interval (0ms, 30ms, 60ms, or 120ms) between writing data to \SiMRA{} rows and executing \SiMRA{}. If the interference is related to charge leakage over time, shorter time intervals should result in weaker interference.} We follow the same setup as \(p_c = 0.0\) and \(p_c = 1.0\) in Figure~\ref{fig:row} for adjacent-row interference (left plot) and Figure~\ref{fig:col_fast} for inter-column interference (right plot), and test 8-row activation on one DRAM chip from each of the three module types.
Figure~\ref{fig:ref} shows the mean Norm.~\({p}_{o1}\) (y-axis) as a function of the waiting time before \SiMRA{} execution (x-axis). Across all conditions, the mean Norm.~\({p}_{o1}\) remains approximately constant regardless of the waiting time (e.g., 0.98 at \(p_c = 0.0\) and 1.02 at \(p_c = 1.0\) in the left plot; 0.92 at \(p_c = 0.0\) and 1.02 at \(p_c = 1.0\) in the right plot), \dtcr{5}{showing that refreshing data more frequently before \SiMRA{} execution does \emph{not} mitigate PuDGhost.}

\footnotetext{We believe PuDGhost would manifest in any \SiMRA{}-capable chip regardless of manufacturer or DRAM \dtcr{5}{type}, as \dtcr{5}{PuDGhost is based on} fundamental analog \dtcr{5}{operational} properties \dtcr{5}{of DRAM} (e.g., charge sharing) and DRAM array design (e.g., hierarchical row decoders, open-bitline architecture) shared across manufacturers and \dtcr{5}{DRAM types}. We \dtcr{5}{already} observe consistent interference patterns across 12 modules with three different die revisions and densities, \dtcr{5}{lending support to} this hypothesis.}

\subsection{Broader Implications}
\subsubsection{Security Concerns for PuD Systems}
PuDGhost exposes two attack vectors in \dtcr{5}{multi-tenant} PuD environments \dtcr{5}{where multiple users share the same DRAM bank}. First, an attacker can (1) allocate memory in DRAM rows adjacent to the victim's \SiMRA{} rows and (2) write specific patterns to \dtcr{5}{these rows} to bias the victim's PuD computation results. Second, conversely, \dtcr{5}{the} attacker could infer statistical properties of \dtcr{5}{the victim's} adjacent-row data by observing bias patterns in \dtcr{5}{the attacker's own} computation results.

\subsubsection{Other \SiMRA{}-based Operations}
PuDGhost can degrade the reliability of a broad spectrum of \SiMRA{}-based operations. For example, \SiMRA{}-based true random number generation (TRNG)~\cite{olgun2021quac,yukel2025simratrng,mutlu2024memory}, \dtcr{5}{\SiMRA{}-based physically unclonable functions (PUFs)}~\cite{baser2026puf}, and other potential future \SiMRA{}-based operations could also be affected. In \S\ref{sec:cas:trn}, we demonstrate that PuDGhost can significantly degrade the randomness quality of \SiMRA{}-based TRNG. This implies that PuDGhost is a critical consideration for TRNG and cryptographic applications that rely on it.

\subsubsection{PuD Simulation Frameworks}
Our characterization reveals that PuD computations performed concurrently under the same \SiMRA{} operation are \emph{not} independent: a given column's \SiMRA{} output depends on data stored in columns that concurrently perform computations. 
This finding has critical implications for PuD simulation frameworks.
Prior PuD simulation models~\cite{de2026count2multiply, li2025pim} assume that errors occur independently in each column. Since whether \dtcr{5}{or not} a column produces a \dtcr{5}{computation} error also depends on the inputs of concurrent computations, these models may not accurately capture application-level output quality.
\section{PuDGhost-Aware Error Mitigation Techniques}\label{sec:mit}
As discussed in \S\ref{sec:rel}, PuDGhost poses a fundamental reliability challenge for PuD systems. Since PuDGhost arises from inherent analog \dtcr{5}{operational} properties \dtcr{5}{of DRAM} and DRAM array design, fully eliminating PuDGhost would be difficult. One promising direction is to design DRAM from the ground up for reliable PuD.
However, such fundamental redesign is not the only path. Prior characterization studies~\cite{kim2014flipping,luo2023rowpress, kim2020revisiting, mutlu2023fundamentally,luo2025revisitingreaddisturbance} of DRAM reliability challenges have led to practical mitigations~\refthrCite{} without requiring DRAM redesign. 

In this section, we discuss mitigation strategies across multiple layers of the PuD computing stack. In \S\ref{sec:mit:cel}, we discuss potential DRAM cell array-level modifications as directions for future research. In \S\ref{sec:mit:prv}, we discuss the relationship between process variation of DRAM circuitry and PuDGhost.
We propose two mitigation techniques for  PuDGhost: 1) new column screening methods (\S\ref{sec:mit:scr}), and 2) a new interference-aware compute row layout (\S\ref{sec:mit:grp}). 
In \S\ref{sec:mit:pat}, we discuss how practical PuD systems can likely be enabled through complementary approaches: 1) PuD ECC schemes and 2) error-tolerant applications. 


\subsection{Cell Array-Level Modifications}\label{sec:mit:cel}
One approach to mitigating PuDGhost is to add modifications to the DRAM cell array to reduce interference during \SiMRA{}. We discuss three potential directions for future research. Each approach requires modifications to the DRAM array, DRAM interface, or memory controller, and occupies a distinct point in the design tradeoff space of reliability, area overhead, and performance. We leave comprehensive evaluation and comparison of these techniques for future work.

\noindent\textbf{Isolation Transistors.} Inserting isolation transistors~\cite{lee2013tiered,chang2016low,luo2020clr} between compute rows and storage rows could electrically disconnect these regions before \SiMRA{} execution, reducing
interference from non-activated rows. However, modern DRAM is implemented with high density~\cite{mutlu2013memory, marazzi2024hifi, nam2024dramscope,mutlu2015main}, and adding transistors \dtcr{5}{incurs} area overhead, potentially \dtcr{5}{eating into} PuD's advantage of throughput per unit area.

\noindent\textbf{Staged Activation.} Our characterization shows that the impact of PuDGhost strengthens with the number of simultaneously activated rows. This motivates exploring staged activation~\cite{prada}, where rows are activated in smaller groups sequentially with sensing performed only after the final group, potentially reducing the impact of PuDGhost while preserving exact PuD computations.
While the feasibility of mitigating PuDGhost requires further investigation, staged activation could offer a useful knob for trading throughput for reliability.

\noindent\textbf{Reference Voltage Adjustment.} Adjusting precharge or reference voltages~\cite{xin2020elp2im} is a promising direction to compensate for the shift in bitline voltage caused by PuDGhost. For example, if adjacent rows are expected to store mostly logic-0 during execution, lowering the reference voltage could compensate for the resulting bias toward logic-0. \dtcr{5}{Performing} such adjustments at runtime to account for the dynamically varying direction and magnitude of PuDGhost could entail significant circuit complexity and power overhead.

\subsection{Reducing Process Variation}\label{sec:mit:prv}
Prior work~\cite{seshadri2017ambit, yuksel2024simultaneous,hajinazar2021simdram,yuksel2023pulsar} demonstrates that process variation in DRAM circuitry can induce PuD computation errors in SPICE simulations. While reducing process variation can alleviate some sources of PuD computation errors, it alone might be insufficient to eliminate them. PuDGhost arises from data-dependent interference from non-operand data, which is \dtcr{5}{fundamentally different from errors caused by process variation}. The interference itself would persist even if process variation were entirely eliminated, potentially still causing PuD computation errors.

\subsection{Interference-Aware Column Screening}\label{sec:mit:scr}
\subsubsection{Naive Column Screening}\label{sec:mit:scr:fail}
A fundamental approach to handling PuD computation errors is to use only identified reliable columns during PuD execution, or to use only DRAM modules where all columns qualify as reliable~\cite{seshadri2017ambit, kubo2025pudtune,hajinazar2021simdram,kubo2025mvdram}. 
A prior study~\cite{kubo2025pudtune} proposes post-manufacturing screening using large sets of random inputs for \SiMRA{} to identify and filter out unreliable columns.
Our experimental findings suggest that screening that does not account for PuDGhost can fail to ensure the reliability of screened columns during PuD execution. The key challenge is that PuDGhost-induced interference is \dtcr{5}{\emph{dynamic}}: non-operand data in adjacent rows can change during PuD execution, and therefore columns labeled as reliable during screening can later produce errors, potentially causing significant application-level errors.
\dtcr{5}{Since column screening with random inputs naturally varies the data in concurrently computing columns across samples of inputs, sufficiently large sets of random inputs can, in principle, capture the effect of interference from concurrently computing columns.}
However, PuDGhost suggests that column screening that does not account for \dtcr{5}{data in adjacent rows} can fail to ensure the reliability of screened columns during PuD execution (see \S\ref{sec:cas:cse}).

\subsubsection{Robust Column Screening and System-Level Support}
To address this challenge, we propose two PuDGhost-aware column screening methods, CS-1 and CS-2.
\begin{itemize}[leftmargin=*]
\item \textbf{(CS-1) Vary data in adjacent rows during screening.} This approach screens columns under multiple data patterns in adjacent rows and retains only columns that are labeled reliable across all tested patterns. It requires no runtime support to control data in adjacent rows during PuD execution, but tends to reduce the number of usable columns because columns must pass screening under all tested patterns.
\item \textbf{(CS-2) Fix data in adjacent rows from screening through execution.} This approach uses the same fixed data pattern in adjacent rows during both screening and PuD execution. Since columns are screened and operated under a single consistent data pattern in adjacent rows, it can preserve more usable columns than CS-1 (see \S\ref{sec:cas:cse}).
It requires runtime support to maintain fixed data in adjacent rows during PuD execution, as well as a compute row layout that ensures adjacent rows remain fixed, as discussed in \S\ref{sec:mit:grp}.
\end{itemize}
We provide a quantitative evaluation of these two column screening methods using real DRAM chips in \S\ref{sec:cas:cse}.

\subsection{Interference-Aware Compute Row Layout}\label{sec:mit:grp}
Mitigating PuDGhost requires careful physical arrangement of compute rows.
In Ambit~\cite{seshadri2017ambit, seshadri2015fast,seshadri2019dram} and its successors~\cite{hajinazar2021simdram, oliveira2024mimdram, oliveira2025proteus}, a small number of compute rows (e.g., six rows) are reserved for \MAJ{3} operations, and the row decoder is modified to simultaneously activate certain triplets of these rows to perform \MAJ{3}. The physical arrangement of these compute rows determines which rows become adjacent to them, directly affecting the requirements for column screening methods. 
With CS-1, since screening tests columns under varying data patterns in adjacent rows, any layout is acceptable. In contrast, CS-2 relies on fixed data in adjacent rows during both screening and PuD execution, making certain \dtcr{5}{compute row} layouts incompatible with CS-2.

\begin{figure}[h]
  \centering
  \includegraphics[width=0.85\columnwidth]{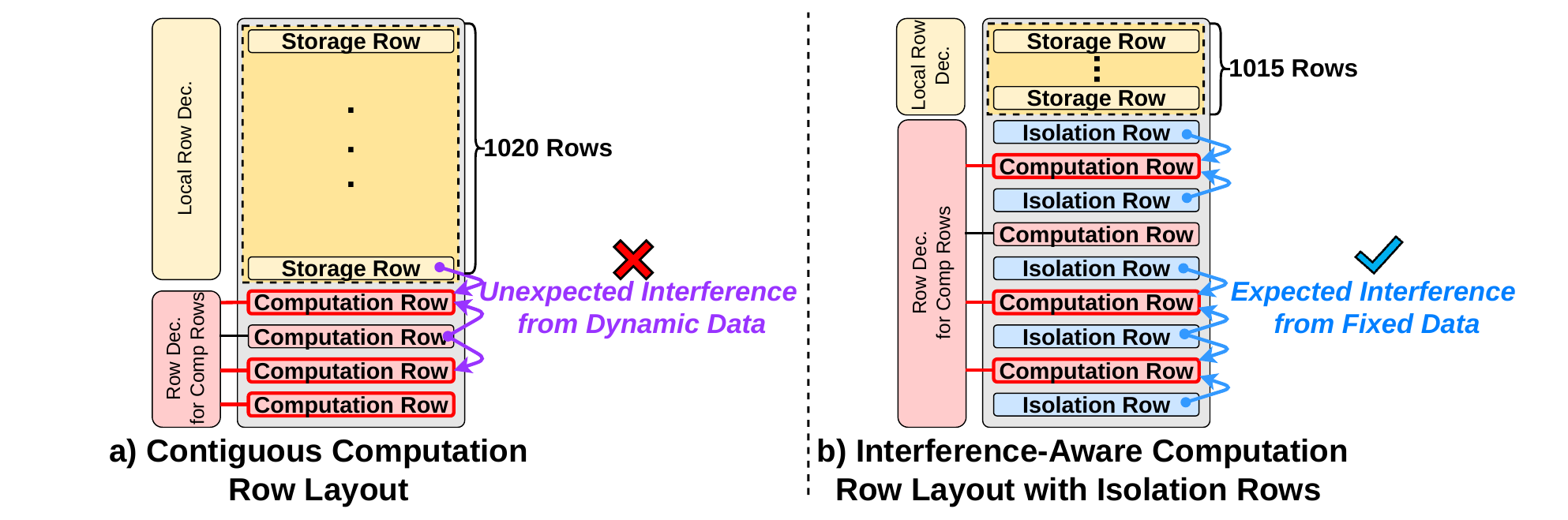}
    \caption{Compute row layouts \dtcr{5}{in a 1024-row subarray with four compute rows}: (a) contiguous layout, (b) isolation-row layout ensuring fixed adjacent-row data.}
  \label{fig:mit-grp}
\end{figure}

\noindent\textbf{Problem with Contiguous Layouts.} 
A contiguous layout (Figure~\ref{fig:mit-grp}a), where all compute rows are placed adjacent to each other, is incompatible with CS-2. In this layout, compute rows participating in one \MAJX{} execution are adjacent to compute rows that can participate in other \MAJX{} executions. Because these adjacent compute rows are overwritten with new operands in other \MAJX{} executions, \dtcr{5}{the fixed data pattern required by CS-2 cannot be maintained.}

\noindent\textbf{Interleaved Layout with Isolation Rows.} 
We propose a compute row layout where each compute row group is \dtcr{5}{placed between} rows with fixed data patterns, called \emph{isolation rows} (Figure~\ref{fig:mit-grp}b). Isolation rows are initialized once with a fixed data pattern and refreshed normally, ensuring that the rows adjacent to compute rows store fixed data that does \dtcr{5}{\emph{not}} change during PuD execution.
This layout introduces no additional latency beyond the one-time initialization. The area overhead is also low: following Ambit with six compute rows, this layout requires seven isolation rows, adding only 0.68\% row-count overhead per 1024-row subarray. Combined with CS-2 column screening, this layout effectively reduces the impact of PuDGhost (as we demonstrate using real \dtcr{5}{DRAM} chips in \S\ref{sec:cas:cse}).

\subsection{Path to Practical PuD Systems}\label{sec:mit:pat}
Our solutions significantly reduce the PuD computation errors (see \S\ref{sec:cas:cse}). To further reduce residual errors toward practical deployment, PuD systems can combine our solutions with two complementary approaches:
1) PuD ECC schemes and 2) error-tolerant
applications.
First, PuD ECC research is advancing~\cite{de2026count2multiply, li2025pim}. For instance, Count2Multiply~\cite{de2026count2multiply} proposes a lightweight ECC scheme that detects and corrects bit errors in PuD computations. Such ECC schemes for PuD can handle residual errors that remain after applying our mitigations.
Second, several important PuD applications, such as deep neural networks (DNNs)~\cite{koppula2019eden} and genomic sequence search~\cite{jahshan2024majork, garzon2026cadm,diab2023framework,alonso2024bimsa,kim2018grim}, have intrinsic error tolerance. For example, EDEN~\cite{koppula2019eden} enables DNN inference on approximate DRAM by retraining models for error resilience. These error-tolerant applications can operate with sufficiently high accuracy even in the presence of some residual errors.



\section{Case Studies}\label{sec:cas}
This section evaluates on real DRAM chips: 1)~our mitigations and their tradeoffs (\S\ref{sec:cas:cse}), and their impact on 2) general matrix-vector multiplication (GEMV) (\S\ref{sec:cas:gem}) and 3) true random number generation (TRNG) (\S\ref{sec:cas:trn}).\footnote{To maintain reasonable testing time, we test three DRAM modules from different chip density and die revision pairs.}

\subsection{Impact of Mitigations on System Tradeoffs}\label{sec:cas:cse}
We evaluate PuD systems equipped with our column screening methods (\S\ref{sec:mit:scr}) and compute row layout (\S\ref{sec:mit:grp}) on real chips, focusing on data patterns and tradeoffs \dtcr{5}{between} reliability, throughput, and capacity overhead.

\noindent\textbf{Experimental Setup.}
Building on the mitigation \dtcr{5}{techniques} in \S\ref{sec:mit:scr} and \S\ref{sec:mit:grp}, \dtcr{5}{following prior work~\cite{kubo2025pudtune}}, we conduct a two-stage experiment targeting \MAJ{3}, the most fundamental PuD primitive.\footnote{\dtcr{5}{Since representative PuD architectures~\cite{seshadri2017ambit,seshadri2015fast,seshadri2019dram,hajinazar2021simdram,oliveira2024mimdram,oliveira2025proteus} do not rely on \Frac{} operations~\cite{gao2022fracdram} to execute \MAJ{3}, we also exclude \Frac{} from our experiments. This ensures that \MAJ{3} errors caused by non-operand data can be attributed to its impact on \SiMRA{}, rather than on \Frac{}.} We use 8-row \SiMRA{}: each of the three operands is duplicated across two rows for redundancy, occupying six rows in total, and the remaining two rows store constant logic-0 and constant logic-1.}
In \textbf{\textit{Stage~1 (screening)}}, each configuration is tested with 8192 samples of random \MAJ{3} inputs, and a column is labeled \emph{reliable} only if it exhibits zero errors across all samples.
In \textbf{\textit{Stage~2 (execution)}}, we target the columns that passed Stage~1 screening, mimicking how a PuD system would operate using only reliable columns. We run 128 additional random-input samples on these columns.
For each \MAJ{3} execution, we (i) write the designated data pattern to adjacent rows, (ii) write random inputs to the \SiMRA{} rows, (iii) execute \SiMRA{}, and (iv) read outputs from the \SiMRA{} rows.
We report three metrics. \textbf{\textit{1) Column Passing Rate (CPR)}} is the fraction of columns that pass Stage~1, which determines the effective throughput of the PuD system. \textbf{\textit{2) Column Break Rate (CBR)}} is the fraction of \dtcr{5}{columns labeled reliable in Stage~1} that produce at least one error in Stage~2. \textbf{\textit{3) Bit Error Rate (BER)}} is the bit error rate of \MAJ{3} in Stage~2 among the columns that passed Stage~1 screening.

\noindent\textbf{Configurations.}
We compare six configurations in three groups, which differ in how data in adjacent rows is managed across Stage~1 and Stage~2.
\textbf{\textit{1) Base}} corresponds to PuDGhost-unaware screening. In Stage~1, all adjacent rows store a single fixed random pattern. In Stage~2, adjacent rows are overwritten before each \MAJ{3} sample with a random pattern whose fraction of logic-1 is uniformly drawn from \(\{0.0, 0.25, 0.50, 0.75, 1.0\}\), simulating unpredictable changes during runtime.
\textit{CS-1} varies the data in adjacent rows during Stage~1 to identify columns that remain reliable across multiple patterns. \textbf{\textit{2) CS-1-01}} divides Stage~1 screening into two halves: all-zeros and all-ones. \textbf{\textit{3) CS-1-sweep}} divides Stage~1 screening into five segments with fraction of logic-1 in \(\{0.0, 0.25, 0.50, 0.75, 1.0\}\). 
\dtcr{5}{We compare these two to test whether screening with only two extreme patterns (CS-1-01) is as effective as a finer-grained sweep (CS-1-sweep).}
Stage~2 uses the same dynamically changing patterns as Base.
\textit{CS-2} fixes the data in adjacent rows to the same pattern in both Stage~1 and Stage~2. \textbf{\textit{4) CS-2-0}} uses all-zeros, \textbf{\textit{5) CS-2-1}} uses all-ones, and \textbf{\textit{6) CS-2-check}} uses a checkerboard pattern.
\dtcr{5}{We test multiple patterns to identify which fixed pattern provides the best tradeoff between reliability and throughput.}

\begin{figure}[h]
    \centering
    \includegraphics[width=\columnwidth]{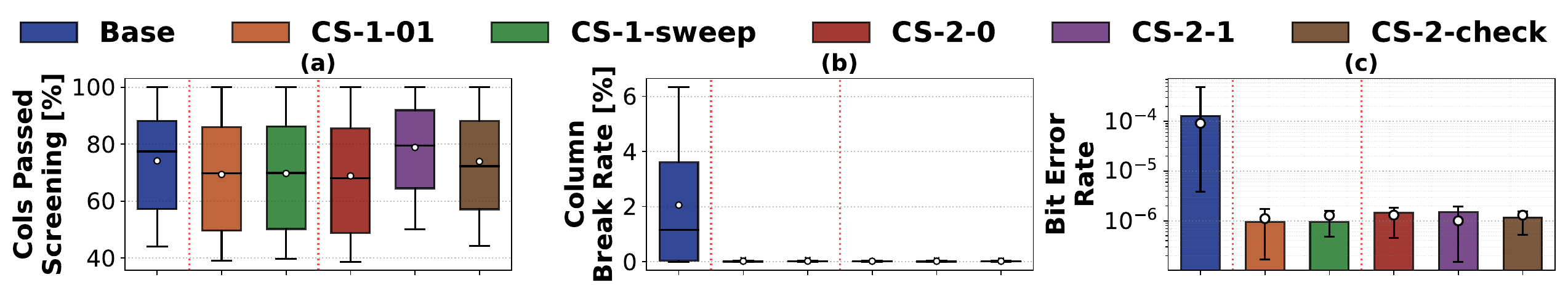}
    \caption{Evaluation of column screening methods.}
    \label{fig:cs_eval}
\end{figure}

\noindent\textbf{Results.}
Figure~\ref{fig:cs_eval} shows the distribution across all tested modules of \dtcr{5}{three metrics} (a)~CPR, (b)~CBR, and (c)~BER\savefootnote{fn:boxplot}{Bar height represents the median, circles indicate the geometric mean, and whiskers show the interquartile range.}
 for the six configurations (x-axis).
Without PuDGhost-aware mitigations (Base), columns that passed screening still exhibit a CBR of 2.1\% and a BER of $9.2\cdot10^{-5}$ on average, demonstrating that PuDGhost-unaware screening fails to filter out unreliable columns.
Our best configuration (CS-2-1) improves CPR (i.e., throughput of PuD execution) by 1.06$\times$ compared to Base, while achieving 125$\times$ lower CBR and 91$\times$ lower BER on average. We estimate that this BER reduction translates to approximately $8.3\cdot10^{3}\times$ lower ECC correction failure rate under the typical configuration of a prior PuD ECC study~\cite{de2026count2multiply}.
For CS-1, we observe no significant difference between CS-1-01 and CS-1-sweep, suggesting that screening with only two extreme patterns (all-zeros and all-ones) is sufficient to identify unreliable columns affected by interference from adjacent rows.
Among CS-2 variants, CS-2-1 achieves the highest CPR, improving it by up to 1.15$\times$ over other CS-2 variants.
Comparing CS-2-1 and CS-1-01, CS-2-1 achieves 1.14$\times$ higher CPR than CS-1-01, at the cost of capacity overhead due to isolation rows (\S\ref{sec:mit:grp}), providing a trade-off between capacity and throughput.

\subsection{Application-Level Study: GEMV Kernel}\label{sec:cas:gem}
We evaluate the application-level impact of PuDGhost and the effectiveness of our mitigations (\S\ref{sec:mit:scr}, \S\ref{sec:mit:grp}) using general matrix-vector multiplication (GEMV), one of the primary target workloads for PuD~\cite{liu2025optipim, kubo2025mvdram, de2026count2multiply, li2025pim,li2024large}.

\noindent\textbf{Experimental Setup.}
We execute GEMV kernels on real DRAM chips using \MAJ{3}, following prior PuD GEMV implementations~\cite{kubo2025mvdram}. We use randomly generated matrices with dimensions $4096 \times N$, where $N \in \{4, 8, 16, 32, 64\}$ at 8-bit precision.\footnote{The performance advantage of PuD over processor-centric GEMV execution becomes more pronounced as \(N\) grows, \dtcr{5}{since a larger \(N\) allows PuD to reduce more data transfer between DRAM and the processor}~\cite{kubo2025mvdram}.}
For each configuration, we use 32,768 columns (i.e., $4096 \times 8$ bits) that passed the respective column screening.


\noindent\textbf{Configurations.}
We compare four configurations (see \S\ref{sec:cas:cse} for detailed screening methodology). \textit{1)~Base-rand} uses the same PuDGhost-unaware configuration as Base in \S\ref{sec:cas:cse}. \textit{2)~Base-worst} represents a PuDGhost-unaware baseline \dtcr{5}{designed to maximize the mismatch between screening and execution}: data in adjacent rows is set to all-ones during Stage~1 but changes to all-zeros during Stage~2. \textit{3)~CS-1} adopts the CS-1-01 configuration. \textit{4)~CS-2} adopts CS-2-1.

\begin{figure}[h]
    \centering
    \includegraphics[width=\columnwidth]{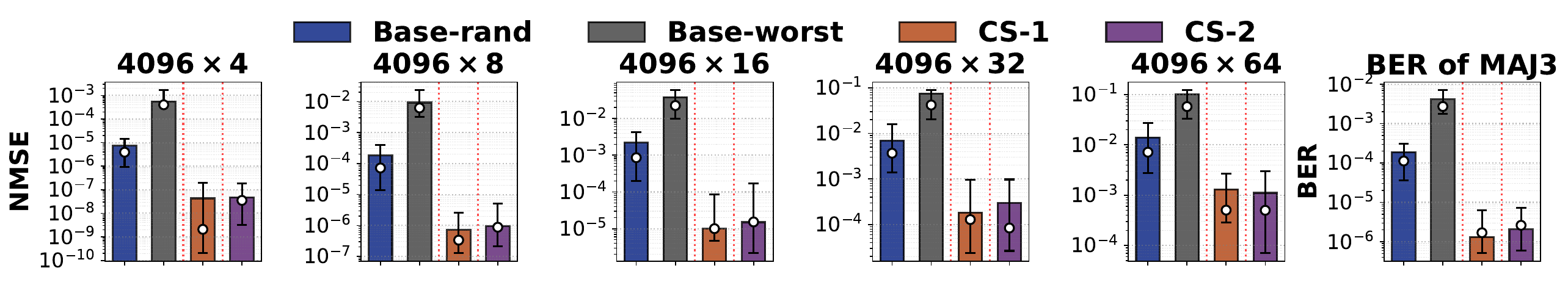}
    \caption{\dtcr{5}{Effect} of PuDGhost and our solutions on GEMV.}
    \label{fig:mvm}
\end{figure}

\noindent\textbf{Results.}
Figure~\ref{fig:mvm}\repeatfootnote{fn:boxplot} shows the normalized mean squared error (NMSE) of GEMV output and the average \MAJ{3} BER for each configuration (x-axis). The first five subplots show NMSE for each matrix dimension ($N \in \{4, 8, 16, 32, 64\}$), and the sixth subplot shows the average \MAJ{3} BER.
NMSE increases with larger $N$ due to longer \MAJ{3} operation chains, which accumulate more errors. Without our mitigations, Base-worst NMSE reaches $2.2\cdot10^{-2}$ at $N=16$ and $5.8\cdot10^{-2}$ at $N=64$, demonstrating that PuDGhost can cause severe degradation in GEMV accuracy.
Our mitigations reduce NMSE below $10^{-3}$ across all tested matrix dimensions. Compared to Base-rand, our mitigations (averaged over CS-1 and CS-2) reduce NMSE by $36\times$ ($N=32$) and $14\times$ ($N=64$), and BER by $55\times$. Compared to Base-worst, the reductions are $413\times$ ($N=32$) and $114\times$ ($N=64$) for NMSE, and $1.3\cdot10^{3}\times$ for BER.

\subsection{PuDGhost's Impact on SiMRA-based TRNG}\label{sec:cas:trn}
We demonstrate the risks of designing \SiMRA{}-based TRNG without accounting for PuDGhost on real DRAM chips.
\SiMRA{}-based TRNG exploits columns that exhibit non-deterministic outputs under fixed \SiMRA{} input~\cite{olgun2021quac,yukel2025simratrng,mutlu2024memory}. These columns, which we refer to as \textit{source columns}, produce outputs that fluctuate due to metastability near the sensing threshold during charge sharing, serving as entropy sources for TRNG.
We investigate how data in adjacent rows and data in non-source columns affect the entropy of source column outputs.

\noindent\textbf{Experimental Setup.}
To measure entropy under each condition, we collect 128 \SiMRA{} output samples with fixed inputs throughout this experiment.
We conduct a two-step experiment. In \textit{Step~1}, following prior work~\cite{yukel2025simratrng}, we search up to 100 candidate input patterns for the \SiMRA{} rows to identify the pattern that maximizes the sum of Shannon entropy~\cite{olgun2021quac,yukel2025simratrng,kim2019drange} across all columns, with data in adjacent rows fixed to a random pattern throughout.
We then identify source columns that exhibit randomness under this pattern.
In \textit{Step~2}, we fix the input pattern found in Step~1 and measure the entropy of source columns while varying data in adjacent rows or non-source columns according to the conditions described below.



\noindent\textbf{Conditions.}
We define seven conditions for Step~2, which differ in what data is changed from Step~1.
\textit{1)~Fixed} is the baseline: data in adjacent rows and non-source columns remains identical to Step~1. This condition corresponds to our mitigation that fixes data in adjacent rows and non-source columns during TRNG operation.
For adjacent-row interference, we modify only the data in adjacent rows from Step~1 while keeping all column data identical to Step~1: \textit{2)~Row-rand} overwrites adjacent rows with freshly generated random data, \textit{3)~Row-0} with all-zeros, and \textit{4)~Row-1} with all-ones.
For inter-column interference, we modify only the data in \SiMRA{} rows of non-source columns from Step~1 while keeping source column inputs and data in adjacent rows identical to Step~1: \textit{5)~Col-rand} sets them to random data, \textit{6)~Col-0} to all-zeros, and \textit{7)~Col-1} to all-ones.

\begin{figure}[h]
    \centering
    \includegraphics[width=\columnwidth]{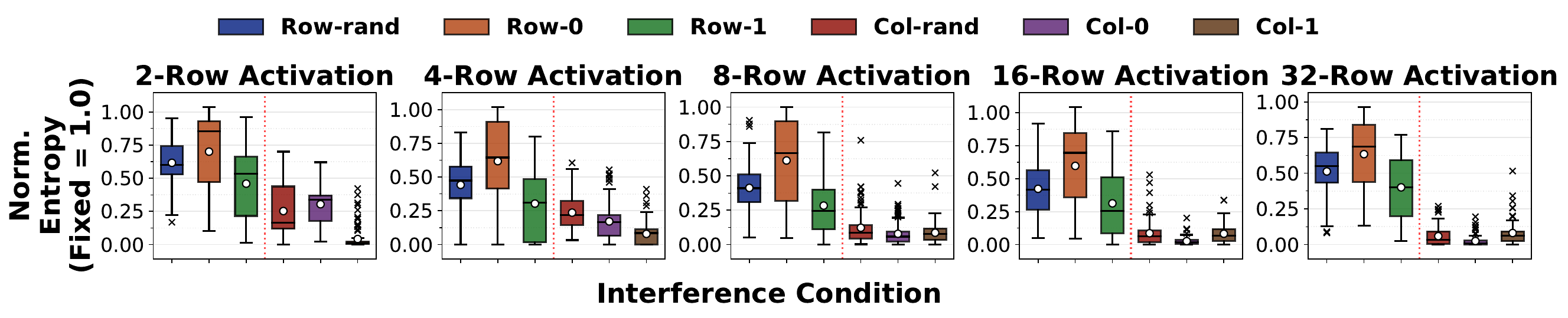}
    \caption{Impact of PuDGhost on \dtcr{5}{the} entropy of \SiMRA{}-based TRNG \dtcr{5}{for different numbers of \SiMRA{} rows}.}
    \label{fig:trng}
\end{figure}

\noindent\textbf{Results.}
Figure~\ref{fig:trng} shows the Norm.~Entropy (y-axis), normalized \dtcr{5}{to} the entropy under the Fixed condition, for the six non-baseline conditions (x-axis), across different numbers of \SiMRA{} rows (each subplot).
Adjacent-row interference reduces entropy compared to Fixed. The worst case is Row-1, which reduces Norm.~Entropy to 0.35 on average across the five activation counts.
Inter-column interference causes even more severe entropy loss. The worst case is Col-1, which reduces Norm.~Entropy to 0.07 on average, meaning that approximately 93\% of the original entropy is lost. Col-0 and Col-rand also drastically reduce entropy to 0.12 and 0.15 on average, respectively.
These results indicate that \SiMRA{}-based TRNG designs that do not account for PuDGhost could unknowingly lose up to 93\% of the entropy that is preserved under Fixed \dtcr{5}{(i.e., our mitigation that maintains the same data in adjacent rows and non-source columns as used during the pattern search)}.

\section{Related Work}\label{sec:related}
\dtcr{5}{To our knowledge, this is the first experimental study demonstrating that non-operand data affects PuD computation results on real DRAM chips.}

\noindent\textbf{DRAM Read Disturbance.}
Many works experimentally characterize read disturbance phenomena on real DRAM chips~\readDisturbCite{}.
RowHammer~\cite{kim2014flipping,kim2020revisiting,orosa2021deeperlook} is the most widely studied example: high-frequency activation of a DRAM row causes bitflips in physically adjacent rows. 
Related phenomena include RowPress~\cite{luo2023rowpress,luo2024combinedrowhammerrowpress,luo2025revisitingreaddisturbance}, which causes \dtcr{5}{bitflips} by keeping rows open for extended periods, PuDHammer~\cite{yuksel2025pudhammer}, which induces bitflips through repeated PuD operations, and ColumnDisturb~\cite{yuksel2025columndisturb}, a bitline-based read disturbance that can induce bitflips in cells sharing the same columns.
\dtcr{5}{Various} works present efficient, wide-ranging attacks and sophisticated mitigations~\isoAllCite{}. 

\noindent\textbf{Multiple-Row Activation-based PuD Operations on COTS DRAM
Chips.}
Prior work demonstrates PuD computations on COTS DRAM using multiple-row activation~\cotsdramCite{}. ComputeDRAM~\cite{gao2019computedram} performs \MAJ{3} and two-input AND and OR operations in DDR3 chips. FracDRAM~\cite{gao2022fracdram} shows that DRAM cells in DDR3 chips can store fractional values. 
\dtcr{5}{DRAM Bender~\cite{safari2022drambender,olgun2023dram} (based on SoftMC~\cite{hassan2017softmc,safari2017softmc}) provides an FPGA-based infrastructure for conducting experiments on real DRAM chips by enabling direct, fine-grained control over DRAM commands.
PiDRAM~\cite{olgun2022pidram,safari2022pidram} provides an FPGA-based framework for end-to-end system evaluation of PuD techniques.}
Prior work~\cite{olgun2021quac,yukel2025simratrng,baser2026puf} executes \SiMRA{} in DDR4 chips to generate true random number \dtcr{5}{and physically unclonable functions}. FC-DRAM~\cite{yuksel2024functionally} demonstrates up to 16-input Boolean operations by performing \SiMRA{} on up to 48 rows in neighboring subarrays. 
PuDTune~\cite{kubo2025pudtune} proposes a methodology to improve the reliability of PuD operations on DDR4 chips using \Frac{} operations. 
While these prior works on PuD using COTS DRAM report errors in PuD operations, none of them investigates interference from non-operand data that is not supposed to participate in PuD computations.

\section{Conclusion}\label{sec:conclusion}
We reveal PuDGhost, a \dtcr{5}{new} interference phenomenon \dtcr{5}{in real DRAM chips} where data in non-activated adjacent rows and columns that concurrently perform computations affects Processing-using-DRAM (PuD) computation results, violating the expectation that each column's computation depends solely on its own operand data. \dtcr{5}{Via rigorous} experimental characterization using \numchips{} real DDR4 DRAM chips, we present \numobsv{} new observations quantifying how non-operand data affects PuD computations. We propose and evaluate solutions \dtcr{5}{to PuDGhost} across multiple layers of the PuD computing stack. We hope our work motivates and guides solutions to enable future PuD systems that are robust.
\section*{Acknowledgments}
We thank the anonymous reviewers of ISCA 2026 for their valuable feedback. We also thank the SAFARI Research Group at ETH Zurich and the members of CASYS at UTokyo for providing a stimulating intellectual environment.
We acknowledge the generous gifts from our industrial partners, including Google, Huawei, Intel, and Microsoft. 
This work is supported in part by the Semiconductor Research Corporation (SRC), the ETH Future Computing Laboratory (EFCL), and the AI Chip Center for Emerging Smart Systems (ACCESS), \dtcr{5}{a Google Security
and Privacy Research Award, the Microsoft Swiss Joint
Research Center}, JST CREST (JPMJCR21D2), JSPS KAKENHI (23H00467), and JST ACT-X (JPMJAX25CC).





\bibliographystyle{unsrt}
\bibliography{refs}

\end{document}